\documentclass[10pt, conference, compsocconf]{IEEEtran}
\usepackage{graphicx}
\usepackage[caption=false]{subfig}
\usepackage{amsmath}
\usepackage{amsfonts}
\usepackage{amssymb}
\usepackage{color}
\usepackage{tabularx}
\usepackage{multirow}
\usepackage{cite}

\usepackage[ruled]{algorithm2e}

\SetAlFnt{\small}
\SetKwRepeat{Do}{do}{while}

\newtheorem{theorem}{Theorem}[section]

\makeatletter  \def\@eqnnum{{\normalsize \normalcolor (\theequation)}}  
\makeatother

\usepackage{setspace}
\usepackage{url}
\usepackage{hyperref}

\ifCLASSINFOpdf
\else
\fi
\begin{document}
\title{The Cloudlet Bazaar Dynamic Markets for the Small Cloud}

\author{\IEEEauthorblockN{Ranjan Pal\IEEEauthorrefmark{1}, Sung-Han Lin\IEEEauthorrefmark{1}, Aditya Ahuja\IEEEauthorrefmark{2}, Leana Golubchik\IEEEauthorrefmark{1}}
\IEEEauthorblockA{\IEEEauthorrefmark{1} University of Southern California \{\emph{rpal, sunghan, leana}\}@usc.edu}
\IEEEauthorblockA{\IEEEauthorrefmark{2} Indian Institute of Technology Delhi \emph{aditya.ahuja}@cse.iitd.ac.in}
}

%\author {
%	\IEEEauthorblockN{Aditya Ahuja\IEEEauthorrefmark{1}, Rajeev Shorey\IEEEauthorrefmark{2}, AuthorName3 %\IEEEauthorrefmark{2}, AuthorName4 \IEEEauthorrefmark{2}}
%	\IEEEauthorblockA{\IEEEauthorrefmark{1} Department of Computer Science and Engineering, Indian %Institute of Technology Delhi}
%	\IEEEauthorblockA{\IEEEauthorrefmark{2} TCS Innovation Labs}
%}


\maketitle

% -*- ispell-local-dictionary: "american"; TeX-master: "cloudshare.tex"; -*-
{\setstretch{1}
\begin{abstract}
The recent emergence of the small cloud (SC), both in concept and in practice, has been driven mainly by issues related to service cost and complexity of commercial cloud providers (e.g., Amazon) employing massive data centers. 
However, the resource inelasticity problem faced by the SCs due to their relatively scarce resources (e.g., virtual machines) might lead to a potential degradation of customer QoS and loss of revenue. A proposed solution to this problem recommends the sharing of resources between competing SCs to alleviate the resource inelasticity issues that might arise \cite{zra}. Based on this idea, a recent effort (\cite{lin2017performance}) proposed SC-Share, a performance-driven static market model for competitive small cloud environments that results in an efficient market equilibrium jointly optimizing customer QoS satisfaction and SC revenue generation. However, an important non-obvious question still remains to be answered, without which SC sharing markets may not be guaranteed to sustain in the long-run - \emph{is it still possible to achieve a stable market efficient state when the supply of SC resources is dynamic in nature and there is a variation of customer demand over time?.} In this paper, we address the problem of efficient market design for SC resource sharing in dynamic environments. We answer our previous question in the affirmative through the use of Arrow and Hurwicz's \emph{disequilibrium process} \cite{ahr1, ahr2} in economics, and the \emph{gradient play} technique in game theory that allows us to iteratively converge upon efficient and stable market equilibria. 
\end{abstract}}


% \fi
\IEEEpeerreviewmaketitle

% -*- ispell-local-dictionary: "american"; TeX-master: "cloudshare.tex"; -*-
\section{Introduction}
Cloud computing is becoming increasingly popular and pervasive in the information technology (IT) marketplace due to its on-demand resource provisioning, high availability, and elasticity. These features allow cloud end users (e.g., individuals, small-scale companies, world-wide enterprises) to access resources in a pay-as-you-go manner and to meet varying demands sans upfront resource commitments \cite{berkeley}. Cloud service providers (Amazon AWS \cite{aws}, Google Compute Engine \cite{computeengine}, and Microsoft Azure \cite{azure}) allow customers to quickly deploy their services without a large initial infrastructure investment. 
%
%Such public cloud providers invest in large-scale data centers, that are typically over-provisioned in order to be able to respond to bursty workloads during peak hours. 

\subsection{The Rise of Small-Scale Data Centers} 
There are some non-trivial concerns in obtaining service from large-scale public clouds, including % loss of data {\em privacy}, 
{\em cost\/} and {\em complexity\/}.
%
%Other concerns include of building services. \ranjan{To expand on these concerns,}
%
Massive cloud environments can be costly and inefficient for some customers (e.g., Blippex \cite{blippex}), thus resulting in more and more customers building their own smaller data centers \cite{cloudservey} for better control of resource usage; for example, it is hard to guarantee network performance in large-scale public clouds due to their multi-tenant environments \cite{mogul2012we}.
%, while smaller-scale cloud providers can offer resources at more competitive prices due to their lower initial investment and maintenance fees, and simpler multiplexing environments; e.g., DigitalOcean \cite{digitalocean} offers higher specification VMs than Amazon EC2 at similar or lower prices.
%
Moreover, smaller data center providers exhibit greater flexibility in customizing services for their users, while large-scale public providers 
%have greater limitations, partly due to their large multi-tenant environments \cite{blippex}
minimize their management overhead by simplifying their services; e.g., Linode \cite{linode} distinguishes itself by providing clients with easier and more flexible service customization.
%
%\ranjan{you have to mention something about privacy with citations here to close the loop in this paragraph. in the other journal paper I submitted on small clouds, you will find some text and citation} 
%Thus, many smaller-scale clouds are available \cite{spotcloud}. \ranjan{This is not the right way to lead to a conclusion, need to reframe the previous sentence.}
%
\emph{The use of small-scale clouds (SCs) is one approach to solve cost and complexity issues.}

Despite the potential emergence of small-scale clouds, the latter due to their moderate sizes, are likely to suffer from resource under-provisioning, thus failing to meet peak demand at times. This leads to a resource provisioning dilemma where the SCs have to make the tradeoff between request loss and the cost of over-provisioning. One way out of this dilemma is for such small clouds to cooperate with each other to help meet each others' user demand via resource sharing at low costs, thereby increasing their individual resources when in need without having to significantly invest in more. Such cooperation is analogous to \emph{Business Clusters} described in mainstream economics which emerge due to, among other factors, shared interests and geographical proximity \cite{bcr}. 
%We note here that since one of the key benefits of SCs is the avoidance of legal and privacy issues, it is apparent that SCs much like Business Clusters would prefer to cooperate with those SCs which lie under one legislative control (modeled in Section II.C as a regulatory agency), thereby alleviating any of their own or their customers' privacy concerns. 
%In addition, some hurdles to centralizing data control in general due to organizational mismatches and lack of incentives have been formalized in \cite{abmd}. 
%Our setting here is a legislatively controlled swarm of networked SDCs that cooperate with each other on resources. Here, cooperation implies paid borrowing of resources

\subsection{Research Motivation}
In this section, we briefly describe the problem setting followed by the challenges that motivate us to alleviate them. 

\textbf{Problem Setting.} The effective sharing or borrowing resources by an SC from its peers involves mutually satisfying the interests of the stakeholders in context. In this paper, we consider three different stakeholders: (i) the SC customers, (ii) profit maximizing autonomous SCs, and (iii) a regulatory agency overseeing certain functioning aspects of the autonomous SCs (e.g., ensuring customer data privacy). The SC customers are interested in achieving certain performance measures for their jobs (e.g., low job response time, cheap storage); the SCs are interested in maximizing revenues obtained from serving customers; and a regulatory agency (e.g., the local government, a federated agency \cite{samaan}\cite{rochr}) is interested in (i) ensuring a proper manner by which the autonomous SCs conduct their business of lending resources to peer SCs (e.g., preserving data privacy, designing policies for (a) customer cost effectiveness by disallowing SCs to charge high customer prices, and (b) maintaining a certain level SC and customer welfare), and (ii) recommending resource exchanges between autonomous SCs in a way (without interfering in the important resource (allocation, scheduling) decisions of the SCs) that encompasses the necessary means for the proper configuration of the resources (e.g., using the \emph{OpenNebula} manager and its external resource lease manager Haizea \cite{smlf}). \emph{We term the above setting as an SC market.}
  
\textbf{The Challenges.} Ideally, an SC would want to service all its customers solely using its own resources. However, the primary barrier to this goal is its individual resource capacity which might not be enough to service peak customer demand. In such a case, the SC can either resort to peer SCs for additional resources, thereby incurring borrowing costs, and/or buy the services of a big public cloud (e.g., Amazon). The latter option is generally more expensive than the former and also likely to be more privacy threatening. Thus, from an SC's viewpoint, its challenge is to satisfy two conflicting objectives: (i) to generate as much revenue by serving its customer demands, and (ii) to incur as low as possible, borrowing and/or buying costs from other clouds. For simplicity purposes, we assume that buying resources from big clouds (e.g., Google, Amazon) is the last resort for an SC in events of low resource availability, and in such events it would try its best to get resources from peer SCs. 
Another challenge is to ensure that at market equilibrium (see below), the SCs and their customers ideally operate on parameters (see Section II) that allow the market to be \emph{efficient}, a condition commonly characterized in microeconomics by certain popular functions (see Section II.C) of market stakeholder utilities, and one that entails optimal social welfare allocation amongst the SCs and their customers. This is a non-trivial and challenging task as the existence of a market equilibrium does not necessarily imply market efficiency \cite{mwg}. \emph{In this regard, the authors in \cite{lin2017performance} show the existence of SC market equilibrium through numerical simulations, and do not provide a general theory for equilibrium existence.} 
In addition to the above mentioned challenges, the SC market is dynamic in nature due to the non-static nature of the supply of SC resources, as well as due to the variations in customer demand over time. This dynamic nature of the SC market is likely to lead to frequent market equilibrium perturbations and potentially a state of market disequilibrium. Conditioned on the achievability of a market efficient equilibrium, a state of eventual disequilibrium will threaten the long-term sustainability of SC markets. Here, the term `market equilibrium' refers to a situation in which all market stakeholders mutually satisfy their interests, in which case an important challenge is to design a stable market that is robust to perturbations and always returns to its equilibrium point(s) when perturbations do occur.

\textbf{Our Goal.} \emph{In this paper, our goal is to formulate the joint `stakeholder satisfaction problem' in dynamic SC environments as an efficient, stable, and sustainable dynamic market/ecosystem design task, and propose an effective solution for it.}

\subsection{Research Contributions}
We make the following research contributions in this paper. 
\begin{itemize}
\item We propose a utility theory based small cloud competitive market model comprising of SC customers, profit maximizing autonomous SCs, and a regulatory agency overseeing some functionality aspects of the SCs, as the market stakeholders. The model mathematically expresses the stakeholder interests in terms of utility functions and paves the path for analyzing SC markets for market equilibrium properties (see Section II). 
\item We analyze our proposed market model via a convex optimization framework for the existence and uniqueness of a static market equilibrium at which (i) the \emph{utilitarian social welfare function} (see Section III for a definition ) is maximized, i.e., the market equilibrium is socially efficient, (ii) the market equilibrium is \emph{Pareto efficient} (see Section III for a definition), (iii) the market is cleared, i.e., the SC supply balances customer demand, and (iv) no stakeholder has any incentive to deviate from the equilibrium. We show that there exists a unique static competitive market equilibrium jointly satisfying (i), (ii), (iii), and (iv), however there are several static market equilibria jointly satisfying (ii), (iii), and (iv). (see Section III). 
\item Using notion of a \emph{disequilibrium process} proposed by Arrow and Hurwicz \cite{ahr1, ahr2}, we apply the \emph{gradient play} technique in game theory \cite{sa} that is based on the theory of differential equations, to investigate the dynamic market setting where a static market equilibrium (conditioned on their existence) is potentially subject to perturbations that might lead to market disequilibrium. In this regard, we show (in theory) that static market equilibrium achieved in small cloud markets is \emph{asymptotically stable} in dynamic market settings. 
%In practice, convergence to static market equilibrium can be achieved iteratively in a distributed manner using methods like the alternative direction method of multipliers (ADMM) \cite{bpcpe}. 
Our use of the gradient play technique is motivated by the fact that in many practical market environments stakeholders (i) find it behaviorally difficult or computationally expensive to play their \emph{best responses} \cite{ft}, (ii) have zero or incomplete knowledge of the utilities of other stakeholders in the market, and (iii) cannot even observe the actions of other stakeholders in the worst case. In such environments, gradient play is a suitable technique to achieve static market equilibrium stability iteratively \cite{flam}, from a state of disequilibrium. More specifically, for our market setting the occurrence of (i)-(iii) is quite likely. Gradient play also works when issues (i)-(iii) do not arise (see Section IV). 
\item We validate our proposed theory through extensive numerical experiments to illustrate the stability of SC markets and the high speed with which such markets converge to stable equilibria despite variations in market supply-demand. Through numerical experiments, we also investigate market equilibria  performance of SC markets with respect to three \emph{Bergson-Samuelson} social welfare functions, viz., the \emph{utilitarian function}, the \emph{egalitarian function}, and the \emph{Rawl's function}, standard in the economics literature \cite{mwg}. Here, we study and compare (i) the fairness amongst market players of a welfare allocation achieved through a given social welfare function and (ii) the amount of social welfare of SC markets under the different social welfare functions, when social welfare optimality is achieved. Via our numerical evaluation, we infer that SC markets are sustainable. (see Section V).  
\end{itemize}
\section{Competitive Market Model}
In this section, we propose a utility theory based small cloud \emph{Walrasian} competitive market model comprising of profit maximizing autonomous SCs, their customers, and a regulatory agency overseeing some functionality aspects of the SCs. A Walrasian competitive market \cite{mwg} represents a \emph{pure exchange economy} without production, where there are a finite number of agents, i.e., SCs in our work, endowed with a finite number of commodities, i.e., computing resources in our work, that gets traded with SC customers and peer SCs. \emph{The aim behind proposing the model is to pave the path for mathematically analyzing SC markets for market equilibrium properties, and derive their practical implications.} 

In this paper, we consider each SC customer to deal with three job types, where each job comprises multiple tasks: (i) Type I jobs that need to be serviced \emph{wholly/entirely} when they arrive (e.g., a user could invoke a regular MapReduce batch job that defines a set of Mappers and Reducers to be executed for the job to complete in its entirety.), (ii) Type II jobs that can be \emph{curtailed} to fewer tasks (e.g., an approximation job like the ones cited in \cite{grass}), the curtailment decision primarily arising from (a) the nature of VM instance prices, (b) the unnecessity of the customer to keep executing a job beyond a certain accuracy already achieved, and (c) the unnecessity of the customer to keep executing the job beyond a certain deadline, and (iii) Type III jobs where certain tasks can be \emph{shifted} over time for future processing, the remaining job tasks requiring service as they arrive (e.g., analyzing a DNA sequence, re-running partially/entirely a current job later when it gets killed in a spot cloud environment due to momentary  unavailability of resources.). Next, we model the stakeholders in the SC market. 
\subsection{Modeling the SCs}  
Let there be $n$ autonomous profit maximizing SCs. Each SC can spread over multiple locations. Customer demand for SC $i$ is a set of processing tasks from its customers (both end-users and peer SCs) that require the use of virtual machines as the primary computing resources. In this regard, we assume that each SC $i$ reserves (allocates) a total of $vm_{i}^{r}$ virtual machines (VMs) in its data center to service demands from its customers. We term such VMs as \emph{reserved} VMs. The value of $vm_{i}^{r}$ is pre-decided by SC $i$ based on the statistics of customer demand patterns observed over a period of time, as one of the factors. \emph{For simplicity, we will focus on VMs representing a single resource type in this paper. A justification is provided in Section 4 of the Appendix \cite{plg}.} In the event that $vm_{i}^{r}$ machines are insufficient to satisfy consumer demands, SC $i$ buys/borrows $vm_{i}^{b}$ VMs from peer SCs. Here, $vm_{i}^{b}$ is the number of \emph{borrowed} VMs available to SC $i$ from its peers. In the event that both reserved and borrowed VMs are insufficient to meet customer demand, SC $i$ resorts to a \emph{public cloud} for $vm_{i}^{pc}$ VM instances. We assume here that a public cloud is large enough to provide any required number of VM instances to SCs. We do not consider communication network bandwidth issues to be a bottleneck to customer service satisfaction in our work. 

Let $c(vm_{i}^{r})$ be the associated operating cost to SC $i$ for reserving $vm_{i}^{r}$ virtual machines to serve its customers. We define $c(vm_{i}^{r})$ via a separable equation of the following form. 
\begin{equation}
c(vm_{i}^{r}) = f_{1}(vm_{i}^{r}) + f_{2}(vm_{i}^{r}),
\end{equation}
where $f_{1}(\cdot)$ (a linear function) and $f_{1}(\cdot)$ (a non-linear function) are functions such that the \emph{marginal operating cost} for SC $i$ is a \emph{general decreasing linear function} of the number of VM instances, i.e., the additional operating cost, $\frac {dc}{dvm_{i}^{r}}$, due to a unit increase in the number of VMs required to service customer demand varies in a negative linear fashion with the number of VMs. We use this type of marginal cost functions in our work due to their popularity in economics  to model diminishing costs/returns \cite{mwg}. We approximate the number of VMs to be a non-discrete quantity. Specifically, for the purpose of analysis, we assume the cost function $c(\cdot)$ to be \emph{concave}, \emph{quadratic}, and \emph{twice continuously differentiable}, i.e., the marginal costs become decreasing linear functions of the number of VM instances. 
We can define one such $c(vm_{i}^{r})$ function as follows. 
\begin{equation}
c(vm_{i}^{r}) = \alpha_{r}^{i}vm_{i}^{r} + \frac{\beta_{r}^{i}}{2}(vm_{i}^{r})^{2},
\end{equation}
where $\alpha_{r}^{i}$ (a positive value) and $\beta_{r}^{i}$ (a negative value) are SC $i$'s cost coefficients for its reserved resources, i.e., virtual machines, such that the marginal operating cost for SC $i$ is a negative linear function. \emph{The above quadratic  form  of  the cost  function, apart from satisfying the property of negative linear marginals, not only allows for tractable analysis,  but  also  serves as  a  good  second-order  approximation for the broader class of concave payoffs} \cite{cbo}. 
We denote $\pi_{i}^{r}$ to be the profit that SC $i$ makes through its reserved VMs for servicing customers, and define the maximum profit that SDC $i$ can make, via the following optimization problem. 
\[\max_{vm_{i}^{r}} \pi_{i}^{r} = \max_{vm_{i}^{r}}[\rho_{i}vm_{i}^{r} - c(vm_{i}^{r})]\]
subject to 
\[vm_{\min_{i}}^{r} \le vm_{i}^{r} \le vm_{\max_{i}}^{r},\]
where $\rho_{i}$ is the per-unit VM instance price charged by SC $i$ to its customers, and $vm_{\min_{i}}^{r}$ and $vm_{\max_{i}}^{r}$ are the lower and upper bounds for the number of VM instances reserved by SC $i$ for its customers. 
We assume that each SC $i$ is small enough not to be able to exert market power over its peer SCs and strategically influence the prices they charge their customers. i.e., each SC is a \emph{price taker} \cite{mwg}. The prices that individual SCs charge their customers are determined by
%either (i) a regulatory agency aiming to achieve a certain optimal social welfare state (see Section II.C) 
individual SCs in price competition with one another in the process of maximizing their own utilities and selling off their endowment. 

Let $c(vm_{i}^{b})$ be the associated operating cost to SC $i$ for borrowing $vm_{i}^{b}$ virtual machines from peer SCs to serve customers, when the reserved VMs are not enough to satisfy customer service demands. Like in the case of formulating $c(vm_{i}^{r})$, we formulate $c(vm_{i}^{b})$ in a manner such that the associated marginal operating costs for borrowing an additional VM instance decreases in a negative linear fashion with the number of VMs. Mathematically, we represent $c(vm_{i}^{b})$ by the following equation. 
\begin{equation}
c(vm_{i}^{b}) = \alpha_{b}^{i}vm_{i}^{b} + \frac{\beta_{b}^{i}}{2}(vm_{i}^{b})^{2},
\end{equation}
where $\alpha_{b}^{i}$ (a positive quantity) and $\beta_{b}^{i}$ ( a negative quantity) are SC $i$'s coefficients for its borrowed virtual machines. We denote by $\pi_{i}^{b}$ the profit that SC $i$ makes when borrowing VMs from peer SCs for servicing customers, and define the maximum profit that SC $i$ can make, via the following optimization problem. 
\[\max_{vm_{i}^{b}} \pi_{i}^{b} = \max_{vm_{i}^{b}}[\rho_{i}vm_{i}^{b} - c(vm_{i}^{b}) - c(vm_{i}^{pc})]\]
subject to
\[vm_{\min_{i}}^{b} \le vm_{i}^{b} \le vm_{\max_{i}}^{b}.\]
Here, (i) $vm_{\min_{i}}^{b}$ and $vm_{\max_{i}}^{b}$ are the lower and upper bounds for the number of VM instances borrowed by SC $i$ for its customers, from peer SCs, (ii) $c(vm_{i}^{pc})$ is the cost to SC $i$ to offload $vm_{i}^{pc}$ VM instances worth of customer demand to a public cloud in the event that $vm_{i}^{r}$ and $vm_{i}^{b}$ VM instances together are not enough to service $i$'s total customer demand. We mathematically represent $c(vm_{i}^{pc})$ in the same manner as $c(vm_{i}^{r})$ and $c(vm_{i}^{b})$, and express it via the following equation. 
\begin{equation}
c(vm_{i}^{pc}) = \alpha_{pc}^{i}vm_{i}^{pc} + \frac{\beta_{pc}^{i}}{2}(vm_{i}^{pc})^{2},
\end{equation}
where $\alpha_{pc}^{i}$ (a positive quantity) and $\beta_{pc}^{i}$ (a negative quantity) are SC $i$'s coefficients for the resources the public cloud uses to service $i$'s offloaded customer demand portions. We do not assume any constraints on the resources available to the public cloud for servicing offloading requests by SCs. 
%Next, we model SC customers. 

\subsection{Modeling SC Customers}
For a customer $j$ who has a Type I job, we express this customer's utility for that job as a concave, quadratic, and twice continuously differentiable separable function, $U_{j}(\cdot)$, defined as follows. 
\begin{equation}
U_{j}(vm_{j}^{e}) = \alpha_{j}^{e}vm_{j}^{e} + \frac{\beta_{j}^{e}}{2}(vm_{j}^{e})^{2},
\end{equation}
where $vm_{j}^{e}$ is the amount of VM instances required to process $j$'s entire job. Similar to the motivation and rationale behind the concave, quadratic cost functions for SCs, the utility function of an SC customer is designed such that the marginal utility for the customer is a \emph{decreasing linear function} of the number of VM instances, i.e., the additional utility increase due to a unit increase in the number of VMs varies in a negative linear fashion with the number of VMs. $\alpha_{j}^{e}$ (a positive quantity) and $\beta_{j}^{e}$ (a negative quantity) in the above equation are $j$'s utility coefficients. 

Like in the case of a customer with a Type I job, for a customer $j$ who has a Type II job, we express his utility for that job as a quadratic twice continuously differentiable function, $U_{j}(\cdot)$, defined as follows. 
\begin{equation}
U_{j}(vm_{j}^{c}) = \alpha_{j}^{e}vm_{j}^{c} + \frac{\beta_{j}^{e}}{2}(vm_{j}^{c})^{2},
\end{equation}
where $vm_{j}^{c}$ is the amount of VM instances required to process $j$'s curtailed job, and is expressed as 
\[vm_{j}^{c} = \kappa_{j}^{1}vm_{j}^{e} + \kappa_{j}^{2}vm_{j}^{e},\,\kappa_{j}^{1}, \kappa_{j}^{2} \in (0,1).\]
Here, $\alpha_{j}^{e}$ (a positive value) and $\beta_{j}^{e}$ (a negative value) are $j$'s utility coefficients for Type I jobs. The interpretation of $vm_{j}^{c}$ is as follows: $ \kappa_{j}^{1}vm_{j}^{e}$ is the number of VMs required to accomplish $j$'s curtailed task, whereas $\kappa_{j}^{2}vm_{j}^{e}$ is the additional number of unused VMs that contribute to $j$'s extra utility when its job is curtailed, and provides it with an overall \emph{perceived} satisfaction greater than that obtained from the utility derived solely using $\kappa_{j}^{1}vm_{j}^{e}$ used VMs for the curtailed job. 

For a customer $j$ who has a Type III job, similar to the case of Type I and Type II jobs, we express his utility for those tasks as a quadratic twice continuously differentiable function, $U_{j}(\cdot)$, defined as follows. 
\begin{equation}
U_{j}(vm_{j}^{s}) = \alpha_{j}^{s}vm_{j}^{s} + \frac{\beta_{j}^{s}}{2}(vm_{j}^{s})^{2},
\end{equation}
where $vm_{j}^{s}$ is the amount of VM instances required to process $j$'s time-shiftable tasks, and $\alpha_{j}^{s}$ ( a positive value) and $\beta_{j}^{s}$ (a negative value) are $j$'s utility coefficients for time-shiftable jobs. 

A customer $j$ can have jobs of all three types. Thus, his aggregate tasks are worth $vm_{j}^{ag} = vm_{j}^{e} + vm_{j}^{c} + vm_{j}^{s}$ VM instances. Therefore, customer $j$'s aggregate utility takes a similar form to his utility for a specific job type, and is given by 
\begin{equation}
U_{j}(vm_{j}^{ag}) = \alpha_{j}^{ag}vm_{j}^{ag} + \frac{\beta_{j}^{ag}}{2}(vm_{j}^{ag})^{2},
\end{equation}
where $\alpha_{j}^{ag}$ (a positive quantity) and $\beta_{j}^{ag}$ (a negative quantity) are $j$'s utility coefficients for his job aggregate. 

We denote $\pi_{j}^{type}$ to be the net utility that customer $j$ generates through getting service for a given job type = $\{e, c, s\}$ from its contracted SC, and define the maximum net utility that customer $j$ can generate, via the following optimization problem. 
\[\max_{vm_{j}^{type}} \pi_{j}^{type} = \max_{vm_{j}^{type}}[U_{j}(vm_{j}^{type}) - \rho_{j}vm_{j}^{type}]\]
subject to
\[vm_{\min_{j}^{type}} \le vm_{j}^{type} \le vm_{\max_{j}^{type}}.\]
Here, $vm_{\min_{j}^{type}}$ and $vm_{\max_{j}^{type}}$ are the lower and upper bounds for the number of VM instances used up by customer $j$'s job type (be it whole, curtailed, shifted, or aggregate). $ \rho_{j}$ is the price paid by customer $j$ to his chosen SC per VM instance used for his job. 
%We now model the regulator. 

\subsection{Modeling the Regulator}
The role of the regulator (e.g., the government, a federated agency) as applicable to our work is to ensure (i) good privacy practices between SC, (ii) the design of policies/mechanisms that enable autonomous SCs to price customers appropriately without making excessive profits through market exploitation, and (iii) an optimum level of social welfare allocation amongst the autonomous SCs at market equilibrium. 
(i) is specific to our problem setting and is one of the most important motivation for the presence of a regulator (see Section I) in the first place\footnote{In practice, using mechanism design theory, the regulator can devise efficient economic mechanisms that enable SCs to find it incentive compatible in protecting the privacy of their customers. However, we do not focus on the design of such mechanisms in this paper.}. 
However, the presence of a regulator brings in other important benefits through (ii) and (iii). (ii) is necessary to prevent any SC from exploiting its customers on service costs. In this work we do not focus on the design of such mechanisms, and assume the existence of one\footnote{Economists Laffont and Tirole have proposed \emph{principal-agent} models in this regard \cite{ltr} which will enable autonomous SCs to charge appropriate prices to customers purely out of self-interest.}, whereas (iii) is important from an economic perspective as \emph{maximizing} social welfare is a key objective in welfare economics because it leads to (a) a certain level of equitability of allocations (in resources or in net utility) amongst the stakeholders, (b) might guarantee \emph{Pareto efficiency} at market equilibrium \cite{mwg}, and (c) an optimal social welfare state denotes the best possible operating point of an economic system. 
A Pareto efficient allocation of utilties amongst a set of stakeholders ensures that at market equilibrium none of the stakeholders can increase their net utility without decreasing any other stakeholder's net utility. The notion of equitability is important in the context of autonomous SC markets because they often operate in a decentralized fashion, and ideally, we would want a social welfare allocation at market equilibrium that does not result in considerable disparity amongst the players' allocations (despite being Pareto efficient). 

In this paper, we define the social welfare function of the regulator to be the sum of the net utilities of the SCs and their customers at market equilibrium. We denote this function by SW, and express it as 
\begin{equation}
SW = \sum_{j\in C}U_{j}(vm_{j}^{ag}) -\sum_{i\in SC}\left(c(vm_{i}^{r}) + c(vm_{i}^{b}) + c(vm_{i}^{pc})\right),
\end{equation}
where $C$ is the set of consumers, $SC$ is the set of small clouds, the first term is the sum of the utility of the consumers, and the second term is the sum of the costs faced by the SCs in $SC$ for servicing customer demands. 
The aforementioned social welfare expression is the standard Bergson-Samuelson \emph{utilitarian social welfare function} in economics \cite{mwg} whose optimality does not focus on equality of resource or utility allocations amongst each class of stakeholders, i.e., the SCs and the customers, but only on Pareto efficiency of resource allocations amongst the stakeholders, and equality of marginal utility allocations amongst the stakeholders. 
Note that due to our autonomous SC setting, the regulator in practice might not have enough say in welfare maximizing resource allocation, and can only expect to have the social welfare function maximized in the best case because it cannot directly enforce optimal strategy choices on the SCs like in a centralized control setting. \emph{The important question here is whether the utilitarian social welfare function is indeed the most appropriate choice for this work.} 

We choose to work with the utilitarian function over two other popular Bergson-Samuelson social welfare functions used in economic applications: the \emph{egalitarian function}, and the \emph{Rawl's function}, for the following reasons:
\begin{itemize}
\item The parameters corresponding to the unique optimal solution of the maximum utilitarian social welfare problem \emph{coincides} with those obtained at the unique equilibrium of a purely distributed market comprising autonomous SC's \emph{without} the presence of a regulator, and are Pareto optimal. This result is due to Arrow-Debreu's first and second fundamental theorems of welfare economics \cite{mwg}. In addition, at market equilibrium, there is equitability in the marginal utilities of all the autonomous SCs (in case of SCs, the utility is represented by cost and is thus a negative utility) and their customers. The parameter coincidence property does not necessarily hold for non-utilitarian social welfare functions.
\item The Rawl's social welfare function focusses on maximizing the minimum resource/utility allocation to any stakeholder (e.g., SC) within the class of market stakeholders. A major drawback of adopting this social welfare function is that it will in general discourage SCs from sharing their resources (even at Pareto optimal system settings) with other SCs (consequently affecting customer QoS satisfaction), thereby challenging the core philosophy behind an SC market, and will not likely be popular with either the SCs or the regulator. A maximin utility allocation among SCs would favor, for example, a regime that reduces every SC to complete ``misery'' if it promotes the well-being of the most ``miserable" SC by even a very small amount. 
\item The egalitarian social welfare function focusses on equalizing the utilities of all market stakeholders in the absolute sense. Similar to the case of Rawl's function, it suffers from the major drawback that it will in general discourage SCs from sharing their resources (even at Pareto optimal system settings) with other SCs. Likewise, it is unlikely to be popular amongst either the regulator or autonomous SCs. For example, if we had to choose between two allocation policies, one under which all SCs would have a cardinal utility of 100, but one SC would have a utility of 99; the second policy under which every SC is ``miserable'' and will have a cardinal utility of 1 unit. The egalitarian regulator would prefer the latter because under this option, every SC has exactly the same utility level.
\end{itemize}

%\emph{We emphasize here that a federated agency can impose centralized control on the SCs (thereby breaking the autonomy of the SCs) to always achieve a socially efficient Pareto optimal allocation amongst stakeholder classes. The challenge in an autonomous SC setting is to achieve the centralized socially efficient Pareto optimal solution in a decentralized manner or getting as close as possible to such a solution. One of our primary goals in this paper is to address this challenge.} 
\section{Static Market Analysis}
In this section we derive and analyze perfectly competitive SC market equilibria. We assume perfect competition amongst SCs due to their lack of economic power in influencing other SCs based on their quantity of VM availability. 
%We assume that the autonomous SC firms decide on their (customer prices, quantity of reserved VMs by SCs) via a price-quantity competition game (e.g., a Cournot game \cite{mwg}). We choose (price, quantity) as the parameters of competition among the SCs due to their publicly observable nature. 
Since prices in perfect competition are \emph{strategic complements} (in the terminology of Bulow, Geanakoplos and Klemperer \cite{bgkr}), i.e., the decrease in an SC's customer price results in the decrease of customer prices charged by other SCs in competition, we are going to eventually converge to a stage where a single uniform customer price will prevail in the SC market \cite{bgkr}. We are interested to know whether such a price results in social welfare optimality. 
%Equivalently, if a federated agency were to centrally impose a customer charging price on all SCs (thereby breaking their autonomy) that would maximize social welfare, what would be the relationship between such a price (quantity) and the market equilibrium price (quantity) outcome of the price-quantity competition game?  
In this regard, we (a) formulate and solve an optimization problem for a regulator who wishes to achieve socially optimal market equilibria that maximizes utilitarian social welfare amongst the market stakeholders, (b) characterize market equilibria in the absence of a regulator and draw comparative relationships between the equilibria obtained, with socially optimal market equilibria. 
%In practice, the competition between SC firms is likely to be imperfect in nature, and Laffont and Tirole have addressed models \cite{lt} under such settings which result in market efficiency. We consider a perfect competition setting in our work for the purpose of simplifying market equilibrium stability analysis in Section 4. 

\noindent \textbf{Optimization Problem Formulation - } Here, we formulate a regulator's optimization problem so as to achieve socially optimal market equilibria. The primary goal of the formulation is to maximize the net utilities for the SC customers, and minimize the net cost of operation of SCs to reach a net maximum social welfare situation amongst the SCs and their customers. We define this problem mathematically as follows:
\[\textbf{OPT:} \quad \max SW\]
subject to 
\[\sum_{j\in C_{i}}vm_{j}^{ag} - (vm_{i}^{r} + vm_{i}^{b} + vm_{i}^{pc}) = 0,\,\forall i\in SC,\]
where the objective function is to maximize social welfare SW (see Equation 9 above) or equivalently to minimize the negative of social welfare (to have a convex objective function to fit the convex programming paradigm), and the constraint is the supply-demand balance equation, with $\sum_{j\in C_{i}}vm_{j}^{ag}$ representing total customer demand, and $(vm_{i}^{r} + vm_{i}^{b} + vm_{i}^{pc})$ representing total SC $i$ supply. $C_{i}$ is the set of customers served by SC $i$. A potential solution to the above optimization problem indicates the parameters at which the SC market can ideally operate and (i) make all stakeholders satisfied to a point that no one has an incentive to deviate, and (ii) maximize the total satisfaction of all the stakeholders together. We denote such an ideal state of market operation as a \emph{static socially efficient market equilibrium.} 

\noindent \textbf{Dual Problem Formulation -} We will solve OPT using the \emph{primal-dual} approach \cite{bv}. The advantage of using the primal-dual approach is that the dual optimization problem of the primal is always convex \cite{bv}, and its solution results in global optima which can be related back to the optimal solution of the primal problem. Before deriving the dual optimization problem, we first define the Lagrangian function of OPT as follows:
\begin{eqnarray*}
L &=& \sum_{i\in SC}\left(c(vm_{i}^{r}) + c(vm_{i}^{b}) + c(vm_{i}^{pc})\right) - \sum_{j\in C}U_{j}(vm_{j}^{ag})\\
&+& \sum_{i\in SC}\rho_{i}\left(\sum_{j\in C_{i}}vm_{j}^{ag} - \rho_{i}(vm_{i}^{r} + vm_{i}^{b} + vm_{i}^{pc}\right),
\end{eqnarray*}
where $\rho = (\rho_{1},....,\rho_{n})$ is the vector of Lagrange multipliers for the constraint in OPT. The dual optimization problem, DOPT, is then defined as follows. 
\[\textbf{DOPT:}\quad \max \inf_{\{vm^{e}, vm^{c}, vm^{s}, vm^{r}, vm^{b}, vm^{pc}, \rho\}} L,\]
 where $vm^{e}, vm^{c}$, and $vm^{s}$ are vectors of customer VM types and $vm^{r}, vm^{b}$, and $vm^{pc}$ are vectors of SC VM types. Note that $vm_{i}^{ag}$ for any customer $i$ equals $vm_{i}^{e} + vm_{i}^{c} + vm_{i}^{s}$. \emph{Thus, the goal here is to find an optimal $\{vm^{e}, vm^{c}, vm^{s}, vm^{r}, vm^{b}, vm^{pc}, \rho\}$ tuple that is an optimal solution to both OPT and its dual.}

\noindent \textbf{Solving the Dual}
The dual optimization problem is convex and its optimal solution is found by applying the \emph{Karush-Kuhn-Tucker} (KKT) conditions \cite{bv} that are stated through equations (10a)-(10g). 
Solving these equations, we obtain the optimal solution to DOPT. Since OPT is convex, applying \emph{Slater's conditions} we obtain strong duality, i.e., a duality gap of zero\cite{bv}, which implies that the optimal solution to OPT coincides with that of DOPT, and there is no loss in the value of the optimal solution by the transformation of the primal problem to its dual. The optimal solution to OPT/DOPT is unique, and is the static market equilibrium. We denote this solution by the tuple $\{vm^{e*}, vm^{c*}, vm^{s*}, vm^{r*}, vm^{b*}, vm^{pc*}, \rho^{*}\}$. We now state the KKT conditions in the form of equations (10a)-(10e) as follows. 
\begin{subequations}
\begin{align}
\frac{d(c(vm_{i}^{r}))}{dvm_{i}^{r}}|vm_{i}^{r*} - \rho_{i}^{*} & = 0,\, \forall i\in SC. \\
\frac{d(c(vm_{i}^{b}))}{dvm_{i}^{b}}|vm_{i}^{b*} - \rho_{i}^{*} & = 0,\, \forall i\in SC.\\
\frac{d(c(vm_{i}^{pc}))}{dvm_{i}^{pc}}|vm_{i}^{pc*} - \rho_{i}^{*} &= 0,\, \forall i\in SC.\\
\rho_{i}^{*} - \frac{\partial(U_{i}(vm_{i}^{e}))}{\partial vm_{i}^{e}}|vm_{i}^{e*} & = 0, \,\forall i\in C. \\
\rho_{i}^{*} - \frac{\partial(U_{i}(vm_{i}^{c}))}{\partial vm_{i}^{c}}|vm_{i}^{c*} & = 0, \,\forall i\in C. \\
\rho_{i}^{*} - \frac{\partial(U_{i}(vm_{i}^{s}))}{\partial vm_{i}^{s}}|vm_{i}^{s*} & = 0, \,\forall i\in C. \\
\sum_{j\in C_{i}}vm_{j}^{ag}(1 - \kappa_{j}^{1} -\kappa_{j}^{2}) & = (vm_{i}^{r} + vm_{i}^{b} + vm_{i}^{pc}),\,\forall i\in SC.
\end{align}
\end{subequations}

\noindent \textbf{Equilibrium in Distributed Autonomous Settings} - The solution to OPT is unique due to the convexity of the dual formulation. \emph{The key question is whether this solution can be realized as a market equilibria in a distributed autonomous setting.} Based on the general equilibrium theory in microeconomics \cite{mwg}, market equilibria in a perfectly competitive autonomous setting of firms is known as \emph{Walrasian equilibria}. It turns out from general equilibrium results in \cite{mwg} that the unique optimal solution to OPT (i) is a competitive Walrasian equilibrium that is Pareto efficient, (ii) satisfies \emph{Arrow-Debreu's} first and second fundamental theorems of welfare economics that establishes the \emph{if and only if} relation between the existence of a Walrasian equilibrium and its Pareto efficiency \cite{mwg}, (iii) maximizes utilitarian social welfare (again derived from Arrow-Debreu's first and second fundamental theorems), and (iv) clears the market by balancing total SC resource supply with consumer and SC resource demand. Thus, in view of points (i) - (iv), a regulator's social welfare maximization objective coincides with the welfare state obtained at market equilibrium in a distributed autonomous firm setting.  \emph{We consider this unique equilibrium state to be the benchmark at which the SC market would be willing to always operate.} However, in practice, for a perfectly competitive market with non-utilitarian social welfare functions, there may be multiple Pareto efficient Walrasian market equilbria that are not socially efficient. 
%In Section V, we compare the social welfare performance of such equilibria induced by the market setting with the optimal social welfare derived from the socially efficient market equilibrium. 

\noindent \textbf{Computing the Socially Optimal Market Equilibrium}
The optimal solution to the dual optimization problem, DOPT, can be obtained in an iterative manner using a gradient approach, the principle behind which is the \emph{Primal-Dual Interior Point Method} \cite{bv}. We adopt the Primal-Dual Interior Point method in our work because it has a polynomial-time complexity to arrive at the optimal solution to convex programs \cite{nn}. The basis of the method is to \emph{progressively} change the argument vector of DOPT so that minima-Lagrange multiplier $\rho$ satisifes the KKT conditions. 

Denote by $v$, DOPT's argument vector sans the Lagrange multiplier $\rho$, $\{vm^{e}, vm^{c}, vm^{s}, vm^{r}, vm^{b}, vm^{pc}\}$.  Applying the Interior Point method to DOPT gives us the the following equations: 
\begin{subequations}
\begin{align}
v(t + \epsilon) & = v(t) - k_{v}\nabla_{x}L\cdot\epsilon.\\
\rho(t + \epsilon) & = \rho(t) + k_{\rho}\nabla_{x}L\cdot\epsilon.
\end{align}
\end{subequations}
Here, $k_{v}$ and $k_{\rho}$ are positive scaling parameters which control the amount of change in the direction of the gradient. Letting $\epsilon \rightarrow 0$, we get 
\begin{subequations}
\begin{align}
\tau_{v} \dot{v}(t) = - \nabla_{v}L,\\
\tau_{v} \dot{\rho}(t) = - \nabla_{\rho}L,
\end{align}
\end{subequations}
where $\tau_{y} = \frac{1}{k_{y}}$ for $y = v, \rho$. The Interior Point Method converges in polynomial time when the duality gap approaches zero, due to the linear and super-linear convergence rate of the method \cite{bv}.

\section{Dynamic SC Markets}
\noindent \textbf{On Dynamic SC Markets} - In practice, an SC market can be dynamic in nature due to the non-static nature of the supply of SC resources and variability of over time of customer demand. This dynamic nature of the SC market is likely to lead to frequent static market equilibrium perturbations, which in turn might (not always) lead to a state of market disequilibrium. Here, the term `disequilibrium' refers to a state when market supply does not equal market demand due to perturbations in market parameters (e.g., customer prices), and as a result all stakeholders do not mutually satisfy their interests. \emph{In such a case, an important challenge is to design a stable market that is robust to perturbations and always returns to its equilibrium point(s) when market disequilibrium results.}
Inspired by the notion of \emph{disequilibrium process} \cite{ahr1}, we propose a \emph{dynamic market mechanism} for SCs. The concept of disequilibrium pertains to a situation where a static market equilibrium is perturbed, potentially to a disequilibrium state, and the underlying players (stakeholders) work together to re-attain the equilibrium. \emph{The main idea behind the disequilibrium process is an iterative sequence of \emph{action} and \emph{state} profiles (see below), i.e., information exchange between the dominant market stakeholders, of VM instance supply and demand levels, and per-unit VM instance prices, to arrive at a desired static equilibrium.} Such an iterative process essentially implies an overall dynamic model with feedback. Our proposed dynamic market mechanism can also be used to re-attain a specific preferred equilibrium point from a given equilibrium point. We first present our dynamic market model and then follow it up with its stability analysis. 

\subsection{Dynamic Model}
Our dynamic model of SC markets consist of a \emph{state space}, $X\subset \mathbb{R}^{n}$, where each state, $\{\rho_{i}\}\in X$, is the profile of per-unit VM instance prices at each SC $i$. The \emph{state dependent payoff}, i.e., profit function for each SC from its reserved resources is given by
\[\pi_{i}^{r} = \rho_{i}vm_{i}^{r} - c(vm_{i}^{r}).\]
The state dependent payoff for each SC from its borrowed resources is given by 
\[\pi_{i}^{b} = \rho_{i}vm_{i}^{b} - c(vm_{i}^{b}).\]
Similarly, state dependent payoff for each SC from resources borrowed from a public cloud is given by 
\[\pi_{i}^{pc} = \rho_{i}vm_{i}^{pc} - c(vm_{i}^{pc}).\]
 The payoff function for the SC customers for a given job type $\in \{e, c, s\}$, is given by 
\[U_{j}(vm_{j}^{type}) - \rho_{j}vm_{j}^{type}.\]
Each SC is assigned a \emph{state dependent action} that permits the SCs and their customers to change their VM instance generation and consumption levels respectively. We assume a \emph{perfect competition} \cite{ft} of VM instance prices amongst the SCs in competition, and following that the action for each SC $i$ consists of commiting a certain amount of VM instances that influences the market-clearing process. In this paper, we use the \emph{gradient play} technique in game theory \cite{sa} to derive the state dependent actions of the SCs and their customers. 
Our use of the gradient play technique is motivated by the fact that in many practical market environments stakeholders (i) find it behaviorally difficult or computationally expensive to play their \emph{best responses} \cite{ft}, (ii) have zero or incomplete knowledge of the utilities of other stakeholders in the market, and (iii) cannot even observe the actions of other stakeholders in the worst case. In such environments, gradient play is a suitable technique to achieve static market equilibrium stability iteratively \cite{flam}. More specifically, for our market setting the occurrence of (i)-(iii) is quite likely. Gradient play also works when issues (i)-(iii) do not arise. \emph{The main idea behind the gradient play technique is the use of ordinary differential equations (ODEs) to describe the path of a perturbed system state to the static market equilibrium state.} Using gradient play, the action for the the $i$th SC is given by 
\begin{subequations}
\begin{align}
\tau_{i}^{r}\dot{vm_{i}^{r}} &= \rho_{i} - \beta_{i}^{r}vm_{i}^{r} - \alpha_{i}^{r}.\\
\tau_{i}^{b}\dot{vm_{i}^{b}} &= \rho_{i} - \beta_{i}^{b}vm_{i}^{b} - \alpha_{i}^{b}.\\
\tau_{i}^{pc}\dot{vm_{i}^{pc}} &= \rho_{i} - \beta_{i}^{pc}vm_{i}^{pc} - \alpha_{i}^{pc}.
\end{align}
\end{subequations}
Here, the parameters $\tau_{i}^{r}$,  $\tau_{i}^{b}$, and  $\tau_{i}^{pc}$ are time constants that describe the speed with which the action of VM instance commitment by SC $i$ can be adjusted, and are free parameters to be determined. 
The goal of SC $i$'s action is to drive the solution $vm_{i}^{r}$, $vm_{i}^{b}$, and $vm_{i}^{pc}$ to $vm_{i}^{r*}$, $vm_{i}^{b*}$, and $vm_{i}^{pc*}$, the solution to Equations 10(a)-10(c) at static market equilibrium. 
It can be seen that the RHSs of 10(a)-10(c) are proportional to the gradient $\nabla_{vm_{i}^{r}}L$, $\nabla_{vm_{i}^{b}}L$, and $\nabla_{vm_{i}^{pc}}L$ respectively, where $L$ is the Lagrangian of OPT. The suite of equations 10(a)-10(c) can be solved independently by SC $i$. 
In a similar fashion, using gradient play, the state dependent action for any SC customer $i\in C$ is given by 
\begin{equation}
\tau_{i}^{ag}\dot{vm_{i}^{ag}} =  \beta_{i}^{ag}vm_{i}^{ag} + \alpha_{i}^{ag} - \rho_{i}.
\end{equation}
$\tau_{i}^{ag}$ is a free parameter to be determined that denotes the speed with which the consumption action of SC customer $i$ can be adjusted.
The goal of the SC customer action here is to drive the solution $vm_{i}^{ag}$ to $vm_{i}^{ag*}$, the solution to Equation 10(d) at static market equilibrium. 
It can be seen that the RHS of 15 is proportional to the gradient $\nabla_{vm_{i}^{ag}}L$, $i\in C$, where $L$ is the Lagrangian of OPT. Equation 15 can be solved independently by each SC customer $i$. 

The dynamics of the pricing mechanism can be expressed via the following equation. 
\begin{equation}
\tau_{\rho_{i}}\dot{\rho_{i}}= \sum_{j\in C_{i}}vm_{j}^{ag}(1 - \kappa_{j}^{1} -\kappa_{j}^{2}) - (vm_{i}^{r} + vm_{i}^{b} + vm_{i}^{pc}),
\end{equation}
where the goal is to drive the solution $\rho_{i},\,\forall i\in SC$ to $\rho_{i}^{*}$, the solution of 10(e) at static market equilibrium. 
Here, $\tau_{\rho_{i}}$ is the free parameter denoting the speed with which $\rho_{i}$ can be adjusted. 
Equations 13-15 represent a dynamic model of the overall SC market. It resembles a repeated negotiation process where SC $i$ responds with a commitment of $vm_{i}^{x}$, $x\in\{r,b,pc\}$ to suggested prices $\rho_{i}$ received from the regulator; SC customer $i$ responds with a consumption amount of $vm_{j}^{type}$, type $\in \{e,c, s\}$, to the same prices. The regulator in turn adjusts its prices to these actions by the SCs and their customers, and returns new prices, $\{\rho_{i}\}$, and the process continues till convergence to the static market equilibrium. 
\emph{A compact representation of the above-mentioned dynamic SC market is presented in Section 2 of the Appendix. This representation paves the way for analytically analyzing the stability of such markets.} 

\subsubsection{A Compact Representation} We need to compactly represent the above dynamic SC market model to pave the way for analytically analyzing the stability of such markets via the \emph{Arrow-Hurwicz} criterion that is based on the theory of Lyapunov stability (see Section IV.B). Using Equations 13-15, our proposed dynamic market mechanism can be compactly represented in the matrix form via the following equation:
\begin{equation}
\begin{bmatrix}
   \dot{ x}_{1}(t)    \\
    \dot{x}_{2}(t)      
\end{bmatrix}
=
\begin{bmatrix}
    A_{1} + \Delta A_{1} & A_{2}  \\
    0 & 0
\end{bmatrix}
\begin{bmatrix}
 	x_{1}(t)    \\
   	x_{2}(t)      
\end{bmatrix}
+
\begin{bmatrix}
 	\bar{\alpha}   \\
   	f_{2}(x_{1} x_{2}).
\end{bmatrix}
\end{equation}
\noindent \emph{Definiton of Equation Parameters.} We now describe the parameters of Equation 16. We have 
\[x_{1}(t) = [VM_{SC}^{r}\, VM_{SC}^{b}\, VM_{SC}^{pc} \, VM_{C}^{e}\, VM_{C}^{c}\, VM_{C}^{s}\,\Delta\, \rho]^{T}\]
that is a vector of dimension $(|SC| + |C| + 2|SC| -1)\times 1$. Here, $|SC| = n$. 
We also have
\[x_{2}(t) = [0]_{n-1 \times 1},\]
and
\[A_{1} = 
\begin{bmatrix}
   -M_{1}  & 0 & 0 & M_{2} \\
    0 & M_{3} & 0 & - M_{4}\\
    0 & 0 & 0 & -M_{5}\\
   -M_{6} & M_{7} & M_{8} & 0
\end{bmatrix}\]
,
\[A_{2} = [ 0\, 0\, -M_{9}\, 0].\]
We define matrices $M_{1}$ to $M_{9}$ as follows:  $M_{1} = Diag(\frac{1}{\tau_{i}^{type}}\beta_{i}^{type})$, type\,$\in\{r, b, pc\}$. We assume that all for a given type, $\tau_{i}^{type}$'s are equal for all $i\in SC$. $M_{2} = Diag(\frac{1}{\tau_{i}^{type}}A_{SC}^{T})$, type\,$\in\{r, b, pc\}$, where $A_{SC} = Diag(1)$. $M_{3} = Diag(\frac{1}{\tau_{i}^{type}}\beta_{i}^{type})$, type\,$\in\{e, c, s\}$. $M_{4} = Diag(\frac{1}{\tau_{i}^{type}}A_{C}^{T})$, type\,$\in\{e, c, s\}$, where $A_{C} = Diag(1)$. $M_{5} = Diag(A'^{T}BA)$, where $A'$ is an $(n) \times (n-1)$ matrix of 1's except for the 0 diagonal elements, $B$ is an $n \times n$ matrix with all entries 1 except for entries of the form $B_{ii}$ that take a value of zero, and $A$ is an $n \times n-1$ matrix.  $M_{6} = Diag(\frac{1}{\tau_{\rho_{i}}^{type}}A_{SC})$, type\,$\in\{r, b, pc\}$. $M_{7} = Diag(\frac{1}{\tau_{\rho_{i}}^{type}}A_{C})$, type\,$\in\{e, c, s\}$. $M_{8} = Diag(\frac{1}{\tau_{\rho_{i}}^{type}}A^{T}BA')$, where $A$ is an $(n-1) \times n$ matrix. $M_{9} = [1]_{n\times n}$. 

The expression $f_{2}(x_{1}, x_{2})$ is a projection function onto the non-negative orthant, and is given by 
\begin{equation}
f_{2}(x_{1}, x_{2}) = [cx_{1} - VM^{\max}]_{x_{2}}^{+},
\end{equation}
where 
$c = BA'R$, $R$ being a rotating matrix. of dimensionality $((|SC| - 1)\times |SC| + |C| + 2|SC| -1)\times 1$, and $VM^{\max}$ denotes a vector of maximum VM instances committed by each individual SC. The $n$th row of the projection $[cx_{1} - VM^{\max}]_{x_{2}}^{+}$ is denoted as 
{\small
\begin{equation} 
    \left[[cx_{1} - VM^{\max}]_{x_{2}}^{+}\right]_{n} = 
\begin{cases}
    \max(0, [cx_{1}]_{n} - VM^{\max}_{n}, & \text{if } [x_{2}]_{n}=0\\
    [cx_{1}]_{n} - VM^{\max}_{n},              & \text{if} [x_{2}]_{n} > 0
\end{cases}
\end{equation}}
$\Delta A_{1}$ in Equation 16 represents the resource availability perturbations due to dynamics of the SC market caused by factors stated in Section IV.A. The value lies in a perturbation set $E$, where $E$ is given by 
\begin{equation}
E = \{\Delta_{A} = \Delta_{SC} - \Delta_{C}|\Delta_{SC} \in E_{SC};\,\Delta_{C}\in E_{C}.\}
\end{equation}
Here,
\[\Delta_{SC} = 
\begin{bmatrix}
   M_{10}  & 0 & 0 & 0 \\
    0 & 0 & 0 & 0\\
    0 & 0 & 0 & 0\\
   M_{11} & 0 & 0 & 0
\end{bmatrix},\]
where matrix $M_{10}$ is given by $Diag\left(\frac{1}{\tau_{i}^{type}}\beta_{i}^{type}(\Delta_{SC})^{2}\right)$, type\,$\in\{r, b, pc\}$, and $\Delta_{SC} = Diag(\Delta_{SC}^{type})$. Matrix $M_{11}$ is given by {\small $Diag\left(\frac{1}{\tau_{\rho_{i}}^{type}}A_{SC}^{T}(I - \Delta_{SC}^{type})\right)$}, and $A_{SC} = Diag(1)$. We also have $E_{SC}$ expressed via the following: 
\[E_{SC} = \{\Delta_{SC}| ||\Delta_{SC}|| = \sqrt{\lambda_{\max}(\Delta_{SC}^{T}\Delta_{SC})} \le \pi_{SC}\},\]
 where $\pi_{SC}$ is a finite constant. Similar to the expression for $\Delta_{SC}$, we have
\[\Delta_{C} = 
\begin{bmatrix}
    0 & 0 & 0 & 0 \\
    0 & 0 & 0 & 0 \\
    0 & 0 & 0 & 0\\
    0 & M_{12} & 0 & 0
\end{bmatrix},\]
where the matrix $M_{12}$ is given by $Diag\left(\frac{1}{\tau_{\rho_{i}}^{type}}A_{C}^{T}(I - \kappa_{j}^{1} - \kappa_{j}^{2})\right)$. We also have 
\[E_{C} = \{\Delta_{C}| ||\Delta_{C}|| = \sqrt{\lambda_{\max}(\Delta_{C}^{T}\Delta_{C})} \le \pi_{C}\},\]
where $\pi_{C}$ is a finite constant. Finally, we express $\bar{b}$ as 
{\small
\[\bar{b} = \left[Diag(\frac{1}{\tau_{i}^{x}}\alpha_{i}^{type}) + Diag(\frac{1}{\tau_{i}^{x}}\alpha_{i}^{x})\Delta_{SC}^{type}\,\, Diag(\frac{1}{\tau_{i}^{y}}\alpha_{i}^{type})\, \,0\right]^{T},\]}
where $x\in\{r, b, pc\}$, and $y\in \{e,c,s\}$. We assume that for given $x, y$, the values of $\alpha_{i}^{x}$ and $\alpha_{i}^{y}$ are equal for all $i$. 

\subsection{Stability Analysis of Dynamic Markets}
In this section, we derive results regarding the stability of static market equilibria in a dynamic SC market setting. Specifically, (i) we derive the dynamic market equilibria obtained through gradient play mechanics and compare it with the socially efficient static market equilibria, and (ii) study the region of attraction around dynamic market equilibria to derive stability connotations. 

\textbf{Case - 1:} We \emph{first} consider stability aspects when $\kappa_{j}^{1}, \kappa_{j}^{2}$ equals zero, i.e., there are no curtailed jobs. In this case, the equilibria of the dynamic SC market described through Equations 10a - 10c (via the use of the gradient play technique), lies in the set
\[E = \{(x_{1}, x_{2})|A_{1}x_{1} + A_{2}x_{2} + \bar{\alpha} = 0\, \cap f_{2}(x_{1}, x_{2}) = 0\}.\]
Let $(x_{1}^{*}, x_{2}^{*})$ be an equilibrium point in set $E$. We then have the following theorem stating the relationship between $(x_{1}^{*}, x_{2}^{*})$ and the unique static SC market equilibrium obtained through Equations 10(a) - 10(e). The proof of the theorem is in the Appendix. 
\begin{theorem}
The equilibrium $(x_{1}^{*}, x_{2}^{*})$ is identical to the unique static market equilibrium obtained from the solution of OPT.
\end{theorem}
\noindent \emph{Theorem Implications.} The theorem suggests that in the absence of curtailed jobs, the equilibrium in a dynamic market setting is unique, and converges to the static market equilibrium in which the market existed initially before it was perturbed. Intuitively, when the SC market is perturbed from its equilibrium setting, a disequilibrium state might result, which will get resolved due to our proposed gradient-play based approach that rolls back the disequilibrium state to the original socially optimal static equilibrium state. In this paper, we are able to roll back to the original state in theory because of our assumptions regarding the nature of utility functions. \emph{In practice, gradient play will guarantee a roll back of a disequilibrium market state to an equilibrium state not necessarily the original equilibrium state from which it was perturbed.} 

We now investigate the stability of the dynamic market equilibrium to find the region of attraction around itself. We introduce a few definitions in this regard. Let $y_{1} = x_{1} - x_{1}^{*}$,  $y_{2} = x_{2} - x_{2}^{*}$. Denote by $V(y_{1}, y_{2})$ a scalar, positive definite Lyapunov function expressed as 
\begin{equation}
V(y_{1}, y_{2}) = y_{1}^{T}y_{1} + y_{2}^{T}P_{2}y_{2},
\end{equation}
where $P_{1}$ and $P_{2}$ are diagonal matrices. We use \emph{Lyapunov functions} from control theory \cite{br} as a standard to prove the stability of an equilibrium of a system represented via ordinary differential equations (ODEs), such as the ones arising in our work in Section IV.A. Let $d$ be expressed as 
\begin{equation}
d = \frac{2\lambda_{min}(P_{2})\psi_{\min}\lambda_{\min}(Q)}{\beta^{2}},
\end{equation}
where $\lambda_{\min}(\cdot)$ denotes the minimum eigenvalue of $Q$, 
\[\beta \ge ||P_{1}A_{2} + R^{T}[1]_{n\times n}P_{2}||_{2},\]
where $R$ is a rotating matrix, and $\psi_{\min} = \min(\psi_{i})$, $\psi_{i}$ being the coefficient of the orthogonal vector $w_{i}$ to express $VM^{\max}$ as $\sum_{i=1}^{n} \psi_{i}w_{i}$. We now have the following theorem characterizing stability of the dynamic market equilibrium. The proof of the theorem is in the Appendix. 
\begin{theorem}
Let $A_{1}$ be Hurwitz. Then the equilibrium $(x_{1}^{*}, x_{2}^{*})$ is asymptotically stable for all initial conditions in 
\[\Omega_{c_{max}} = \{(y_{1}, y_{2})||V(y_{1}, y_{2}) \le c_{max}\}\,\mathrm{for}\, c_{max} > 0, \]
such that 
\[\Omega_{c_{max}} \subsetneq D = \{y_{2} \ge 0|||y_{2}||_{2} \le d\}\]. 
\end{theorem}
\noindent \emph{Theorem Implications.} Intuitively, the theorem states that irrespective of any initial state the market is in, on being perturbed, it will always come back to an equilibrium state from a disequilibrium state. The \emph{Hurwitz} (not the same as Hurwicz) nature of matrix $A_{1}$ is determined from the time constants in Equations 13-15. Most real systems satisfy the Hurwitz criterion in that $A_{1}$ will be a \emph{real square matrix} constructed with coefficients of a real polynomial. 

\textbf{Case 2:} We \emph{now} consider stability aspects when $\kappa_{j}^{1}, \kappa_{j}^{2}$ \emph{does not} equal zero. In this case, the equilibria of the dynamic SC market described through Equations 13(a) - 13(c), also lies  in the set $E$. We define $y_{1}, y_{2}$, and $V(y_{1}, y_{2})$ as before but define $d_{\Delta}$ as 
\begin{equation}
d_{\Delta} = d - d_{\Delta_{SC}} + d_{\Delta_{C}},
\end{equation}
where $d$ is the same as in Equation (21),  $\Delta_{SC}$ and $\Delta_{SC}$ represent the supply demand perturbation matrices, and $d_{\Delta_{SC}}$ and $d_{\Delta_{C}}$ are given by 
\begin{subequations}
\begin{align}
d_{\Delta_{SC}} &= \frac{4\lambda_{min}(P_{2})\psi_{\min}||P_{1}||_{2}\pi_{i}|i\in SC}{\beta^{2}}.\\
d_{\Delta_{C}} &= \frac{4\lambda_{min}(P_{2})\psi_{\min}||P_{1}||_{2}\pi_{j}| j\in C}{\beta^{2}}. 
\end{align}
\end{subequations}
We now have the following theorem characterizing market stability. The proof of the theorem is in the Appendix. 
\begin{theorem}
Let $A_{1}$ be Hurwitz, and let 
\begin{equation}
\pi_{SC} - \pi_{C} < \frac{\lambda_{\min}(Q)}{2||P_{1}||_{2}}
\end{equation}
Then the equilibrium $(x_{1}^{*}, x_{2}^{*})$ is asymptotically stable for all initial conditions in 
\[\Omega_{c_{max}} = \{(y_{1}, y_{2})||V(y_{1}, y_{2}) \le c_{max}\}\,for\, c_{max} > 0, \]
such that $\Omega_{c_{max}} \subsetneq D = \{y_{2} \ge 0|||y_{2}||_{2} \le d_{\Delta}\}$. 
\end{theorem}
\noindent \emph{Theorem Implications.} Similar to the implications of Theorem 4.2, this theorem states that irrespective of any initial state the market is in, on being perturbed, it will always come back to an equilibrium state from a disequilibrium state. 
\section{Numerical Evaluation}
In this section, we numerically evaluate our dynamic market model to investigate (a) static market efficiency and average stakeholder utility under different Samuelson-Bergson welfare metrics, and (b) stability behavior of dynamic markets. In the absence of a theoretical study (not our focus in this paper), (a) is important to get an approximate idea of the gaps in net stakeholder utility achieved via different welfare functions. (b) is important to characterize the speed of convergence of an SC market to go from a state of disequilibrium to a state of equilibrium. The first part of this section describes the evaluation setting, and the second part analyzes the results. 
\subsection{Evaluation Setup}
As a representative numerical evaluation setting, we consider five SCs and 15 customers (not including other SCs). Each SC has five customers each and they are tied to the SCs throughout the entire duration of the experiment. Peer SCs are assumed to be altruistic w.r.t. VM borrowing. The market parameters for the SCs and the customers are shown in Tables 1 and 2 respectively. We simulate a perfect price competition game between the SCs, and use the \emph{tatonnement process} (TP) \cite{btsk} to converge to a static market equilibrium in practice for a distributed setting. \emph{Tatonnement} is a trial-and-error process similar to the \emph{hill climbing} approach in local search theory by which equilibrium is reached in competitive markets via a distributed fashion. 
As a measure of static market efficiency we investigate and compare the utilitarian SW function values at market equilibrium for \emph{utilitarian, egalitarian}, and \emph{Rawlsian} \emph{(see Section II.C for more details)} regulators. \emph{Note that the utilitarian SW function reflects the net stakeholder utility, and out goal is to study the net stakeholder utility at market equilibrium for regulators with different utility equitability mindsets.}  
For the parameter values in Tables 1 and 2, we run numerical evaluations for all possible permutations (instances) of values that are applicable to SCs and their customers, and report the mean value of the results obtained \emph{(with the exception of Figure 1c which reports (without loss of generality) on individual permutations)}. Note that each permutation of values can be considered as a different market setting. 
To experiment on dynamic markets, as a representative example, we fix $\tau_{\rho}$ to be the same for all SCs and vary it in the interval $[0, 5]$. Similarly we fix $\tau^{ag}$ to be the same for all 15 customers and vary it in the interval $[.05, .2]$. We also make $\kappa^{1}$ and $\kappa^{2}$ to be equal for all customers and vary it in the interval $[0, 0.05]$. To provide a rationale behind the values for the time constant, $\tau_{\rho}$, first note that it represents the market time scale for the update of prices. Small values of $\tau_{\rho}$ implies a fast update in real-time price, which can introduce volatility. On the other hand, high values of $\tau_{\rho}$ contributes to reduced volatility. The time constant, $\tau_{ag}$, represents the reciprocal of consumer demand elasticity. 

\begin{table}[h!]
  \begin{center}
    \caption{Parameters of Cost Functions for SCs}
    \label{tab:table1}
    %\hline
    \begin{tabular}{c|c|c|c|c|c}
      SC\# &$VM^{\min}$ & $VM^{\max}$ & $\tau_{SC}$ & $\beta$ & $\alpha$\\
      \hline
      SC1 & 0 & 200 & 0.6 & -0.3 & 90\\
      SC2 & 0 & 200 & 0.2 & -0.6 & 102\\
      SC3 & 0 & 250 & 0.6 & -0.25 & 80\\
      SC4 & 0 & 250 & 0.6 & -0.25 & 80\\
      SC5 & 0 & 200 & 0.2 & - 0.01 & 20\\
     \hline
    \end{tabular}
  \end{center}
\end{table}

\begin{table}[h!]
  \begin{center}
    \caption{Parameters of Utility Functions for Customers}
    \label{tab:table2}
    %\hline
    \begin{tabular}{c|c|c|c|c|c}
      Customer(C)\# &$VM^{\min}$ & $VM^{\max}$ & $\tau_{C}$ & $\beta$ & $\alpha$\\
      \hline
      C1, C2 & 60 & 100 & 0.1 & -0.5 & 168\\
      C3, C4 & 60 & 100 & 0.1 & -0.15 & 140\\
      C5, C6 & 70 & 80 & 0.2 & -0.35 & 140\\
      C7, C8 & 20 & 60 & 0.2 & -0.2 & 100\\
      C9, C10 & 30 & 60 & 0.2 & -0.3 & 120\\
      C11, C12 & 20 & 40 & 0.2 & -0.1 & 125\\
      C13, C14, C15 & 30 & 60 & 0.2 & -0.5 & 135\\
     \hline
    \end{tabular}
  \end{center}
\end{table}

\subsection{Analysis of Evaluation Results}
In the first part of this section, we analyze SC cost and customer utility allocations at market equilibrium under the utilitarian, egalitarian, and Rawlsian SW paradigms. In the second part, we analyze the stability of various dynamic market settings, and also how fast a dynamic market converges to a stable equilibrium. 

\noindent \textbf{Static Market Equilibrium Performance.} Using TP and in the presence of a regulator with different social welfare (SW) mindsets, we arrive at a different single market equilibrium (ME) maximizing SW. Note here that ME might not be unique, and in this case TP will converge locally to a ME in a distributed manner. The regulator will then have the option to work upon the ME to maximize SW.  We observe (as a mean of multiple instances) from Figures 1(a\&b) that with respect to SC and customer allocation ratio equitability, Egalitarian MEs are the best as they ensure nearly identical cost and utility allocation ratios across all autonomous SCs and customers respectively, followed closely by Rawlsian MEs, and utilitarian MEs that are not very fair (equitable) in the utility allocation sense. 
Here, we define \emph{allocation ratio} as the ratio of the cost (utility) of SC (customer) $i$ to the maximum cost (utility) of any SC (customer) at ME, for each given market type. On the other hand, we see that market equilibrium in utilitarian markets, MEs lead to a considerably greater additive stakeholder satisfaction (utility) (see Figure 1c.) when compared to egalitarian and Rawlsian markets., i.e., the utilitarian SW metric is highest in utilitarian markets. This is true from theory as marginal stakeholder utility at utilitarian ME is equal across all stakeholders. In addition, from theory, SW maximizing ME in utilitarian competitive markets are \emph{always} Pareto optimal. In Figure 1c, $U-SW_{OPT, t}$, $t\in \{U, E, R\}$, denotes the utilitarian social welfare value at the optimal market situation of type $t$, and $\frac{U-SW_{OPT, t}}{U-SW_{OPT, U}}$ is the ratio of the utilitarian social welfare value at the optimal market situation of type $t$ to the optimal utilitarian social welfare at utilitarian ME. 

\noindent \textbf{Dynamic Market Stability Performance.} Through Figures 2, 3, and 4, we study dynamic markets for three different instances of $(\kappa_{1}, \kappa_{2})$ pairs, for utilitarian, Rawlsian, and egalitarian market types. For each instance, and for any market type, we observe that low values of $\tau_{\rho}$ for a given instance correspond to market instability, i.e., a state of disequilibrium, because they imply a fast update in SC prices charged to customers, indicating market volatility in supply and demand as well. Here, stability is indicated through the maximum of the eigenvalues of Hurwitz matrix $A_{1}$ (see Section III) formed from the market instance, which are negative in the stable zone, and positive in the unstable zone. It is logical to expect that market instability can be reduced if the price update is slower, i.e., if $\tau_{\rho}$ is larger. In this regard, we observe that the reduction in market instability takes place the slowest for egalitarian market types because in such markets there is a strict requirement of absolute stakeholder utilities to be equal at ME. The reduction in market instability is the fastest for utilitarian market types. 
We also observe from figures 2-4 that market volatility is increased due to a decrease in $\tau^{ag}$ values because the latter trend corresponds to the increase in demand elasticity which contributes to a market being volatile. We infer from the plots that it is possible to design an SC market where volatility (arising due to either low $\tau_{\rho}$ or low $\tau^{ag}$ values) can be contained by increasing market latency, i.e., increasing $\tau_{\rho}$ values. With respect to the speed of convergence, from Figures 2-4, we observe in general that SC markets converge fast to the stable zone, i,e., even at low values of $\tau_{rho}$, but the speed of convergence increases with increasing $\kappa_{1}, \kappa_{2}$ values. This is because increasing $\kappa$ values indicate more demand curtailment by SC customers, potentially leading to non-volatility in supply and prices. 

\begin{figure*}[htb]
\centering
 \begin{tabular}{@{}cc@{}}
  \includegraphics[width=0.33\textwidth]{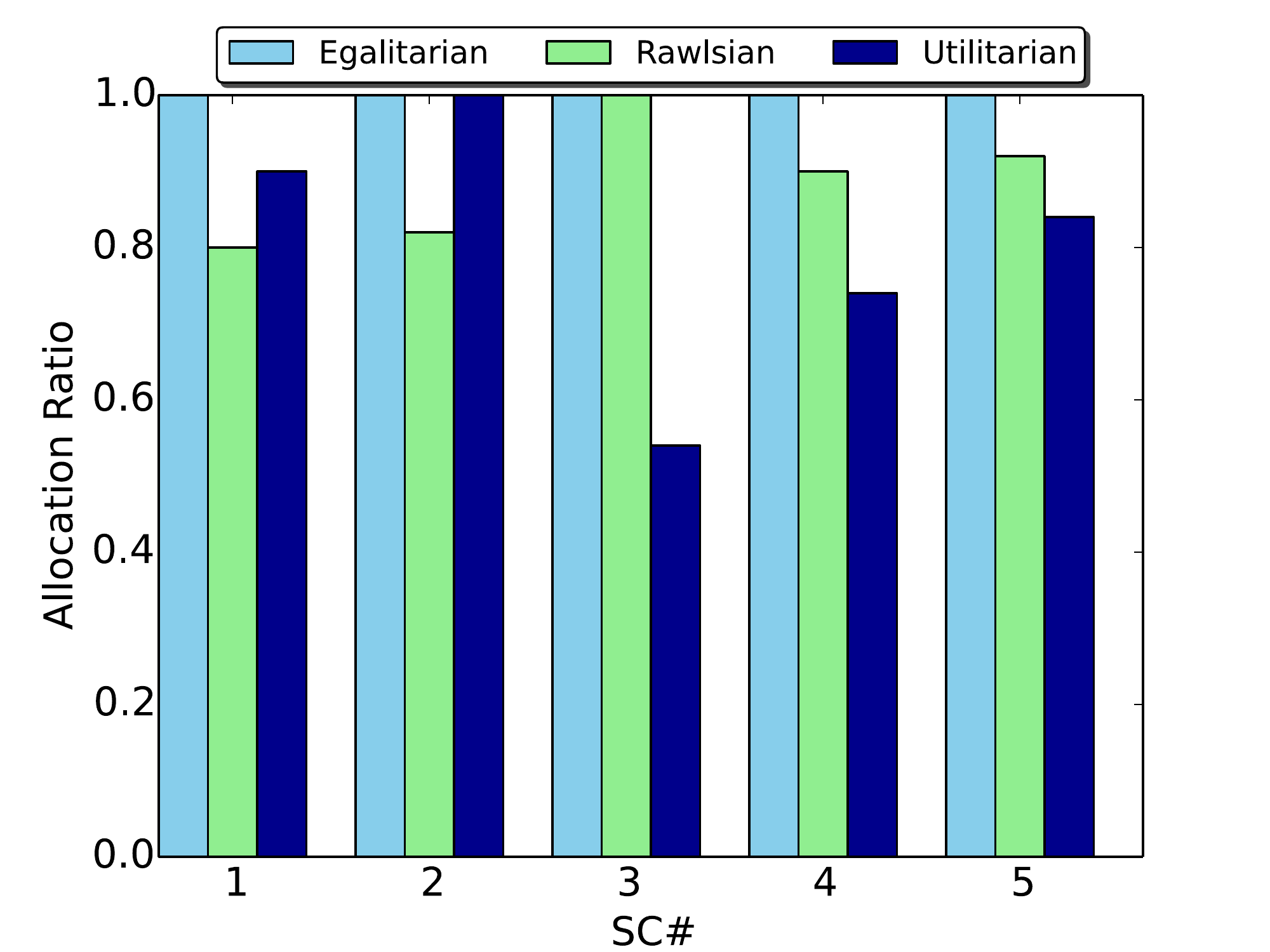} &
  \includegraphics[width=0.33\textwidth]{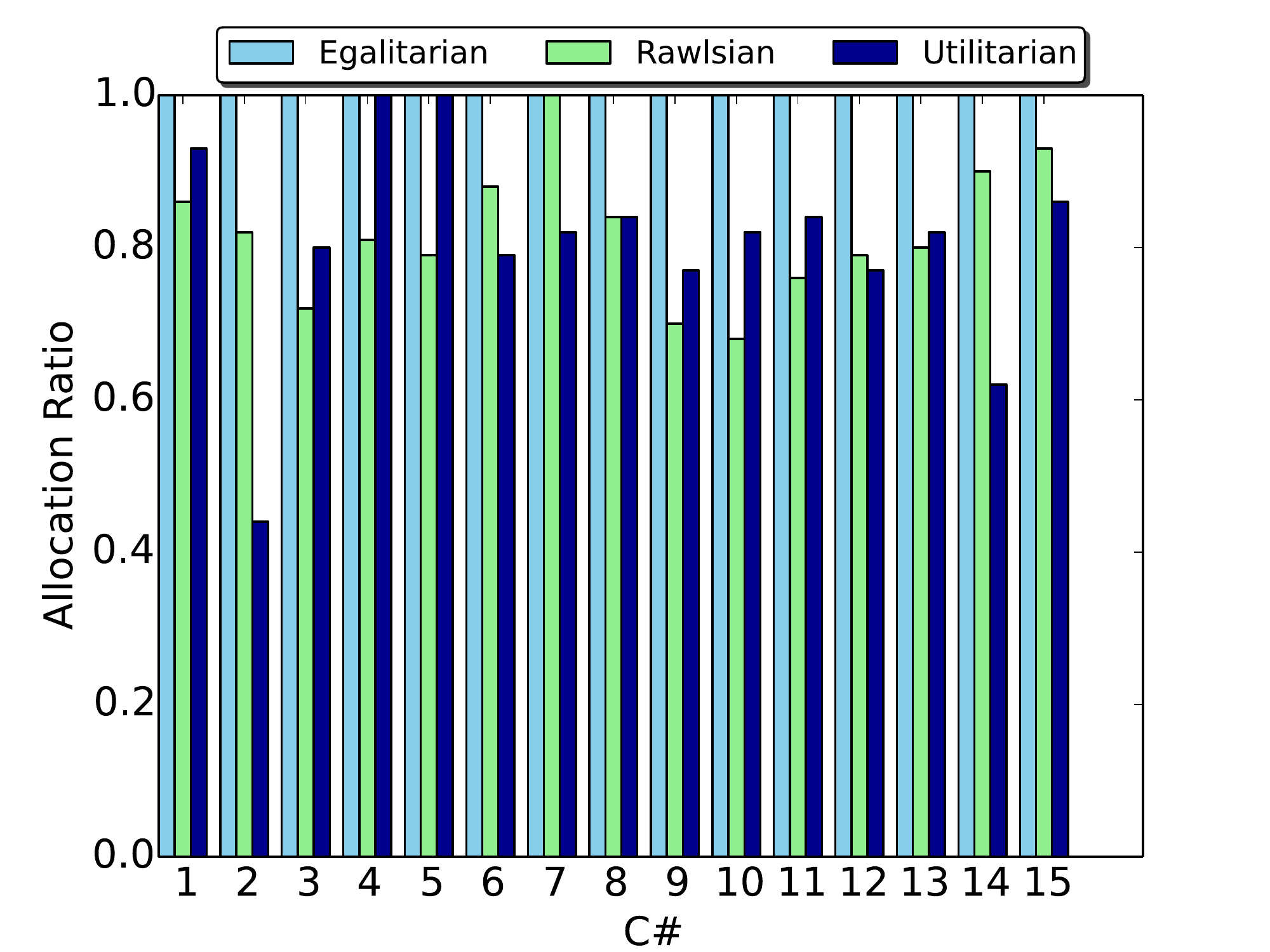}
  \includegraphics[width=0.33\textwidth]{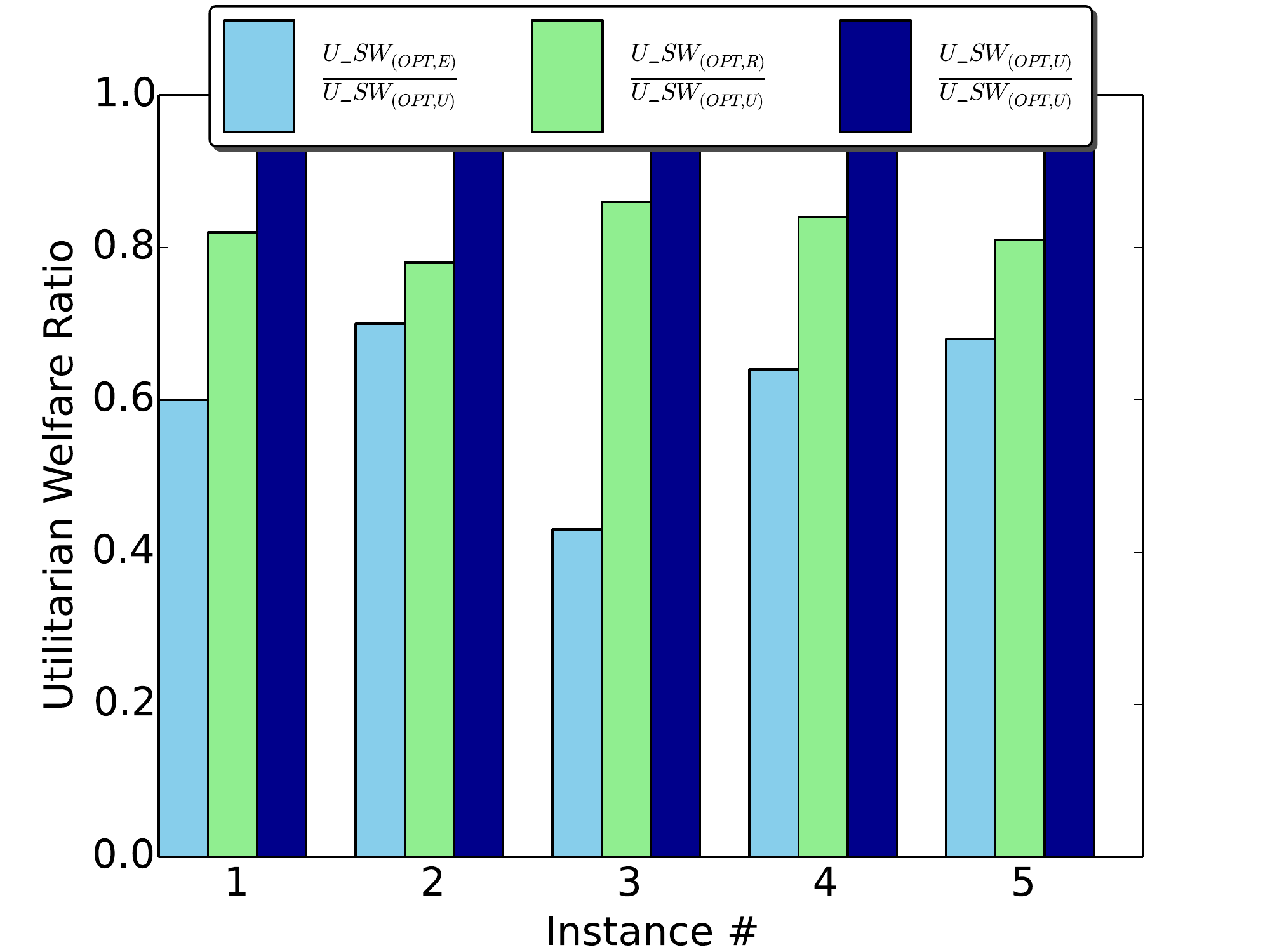} 
\end{tabular}
\caption{ME Performance (a) SC Costs (left), (b) Customer Utilities (middle), (c) Social Welfare Ratio (right) w.r.t to various SW metrics}
\end{figure*}

\begin{figure*}[htb]
\centering
 \begin{tabular}{@{}cc@{}}
  \includegraphics[width=0.33\textwidth]{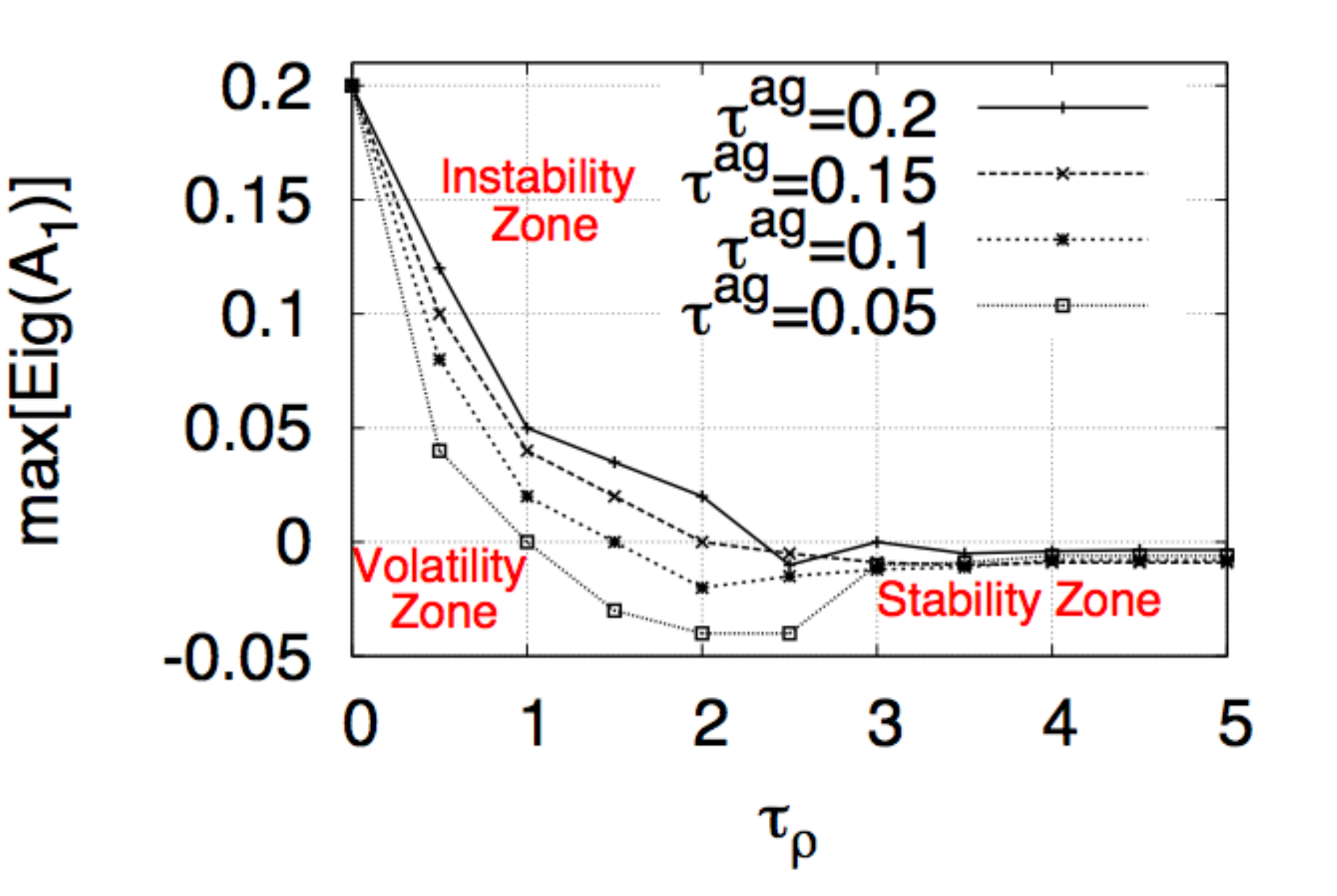} &
  \includegraphics[width=0.33\textwidth]{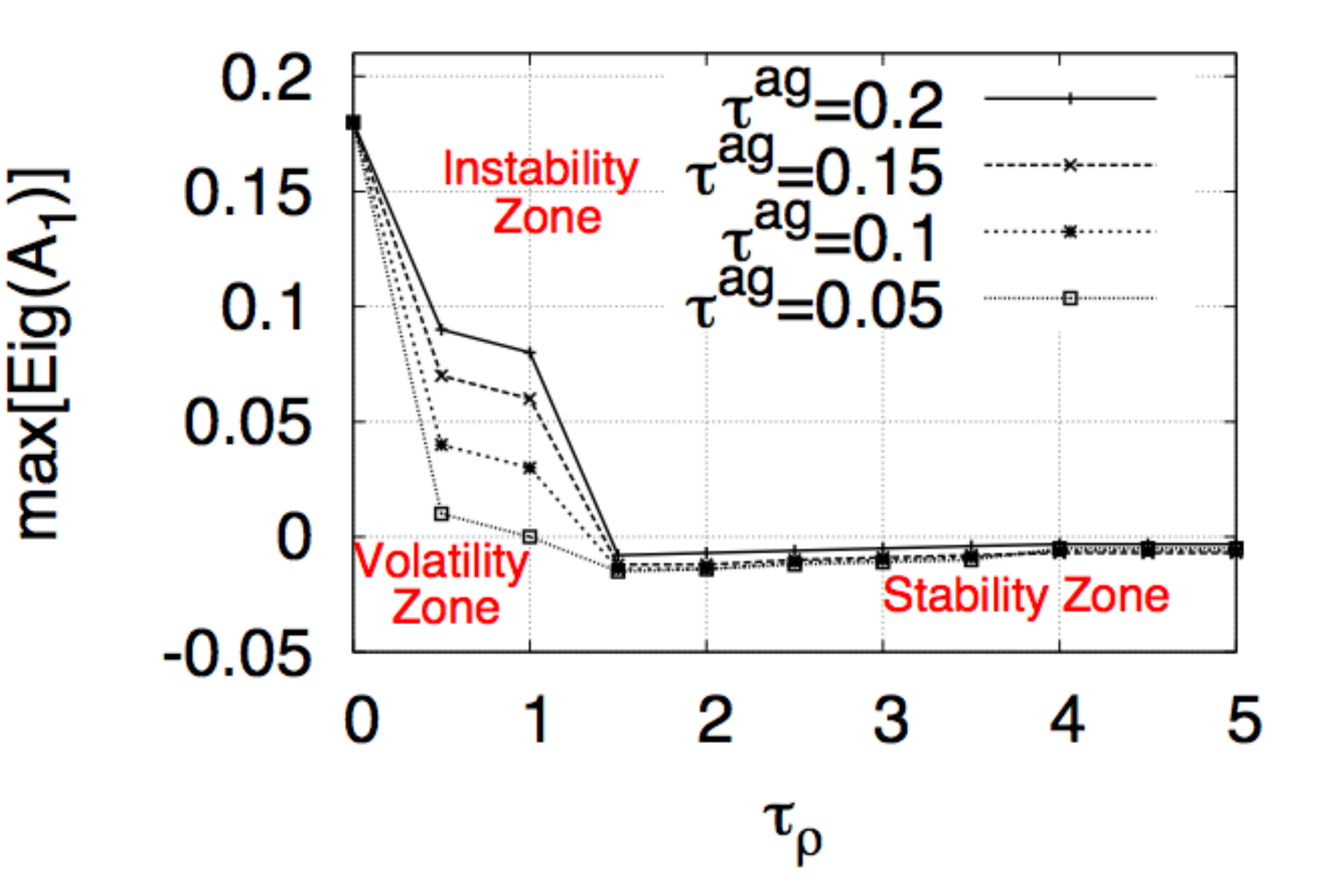}
  \includegraphics[width=0.33\textwidth]{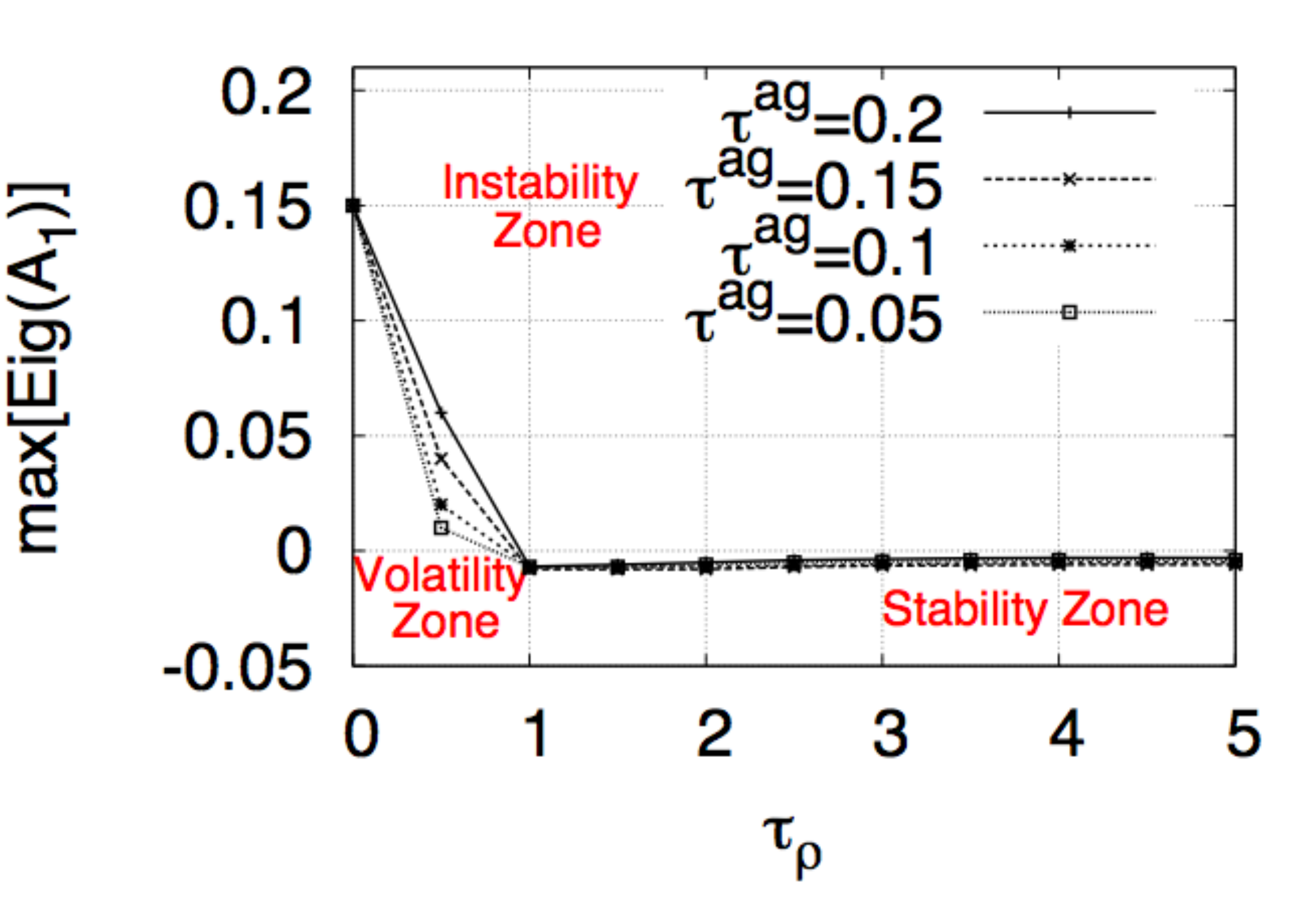} 
\end{tabular}
\caption{Market Stability Performance when (a) $\kappa_{1} = \kappa_{2} =0$ (left), (b) $\kappa_{1} = \kappa_{2} =0.02$ (middle), (c) $\kappa_{1} = \kappa_{2} =0.05$ (right) [Utilitarian]}
\end{figure*}

\begin{figure*}[htb]
\centering
 \begin{tabular}{@{}cc@{}}
  \includegraphics[width=0.33\textwidth]{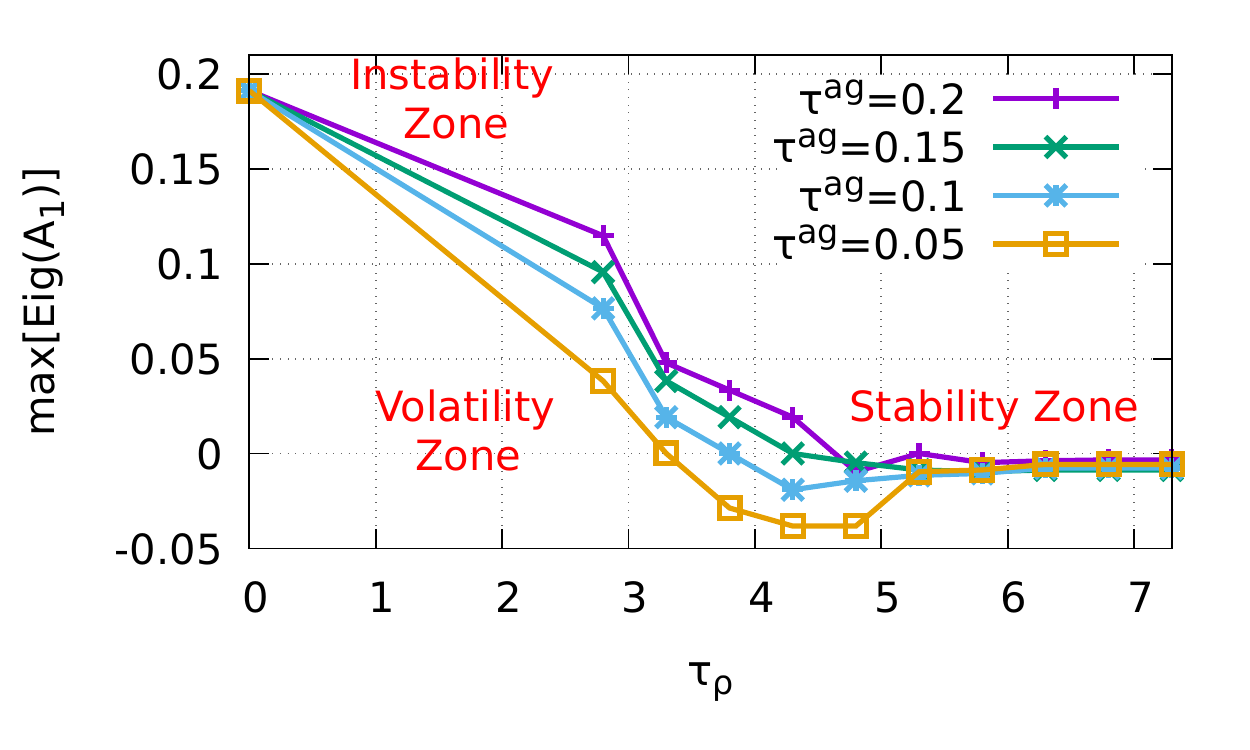} &
  \includegraphics[width=0.33\textwidth]{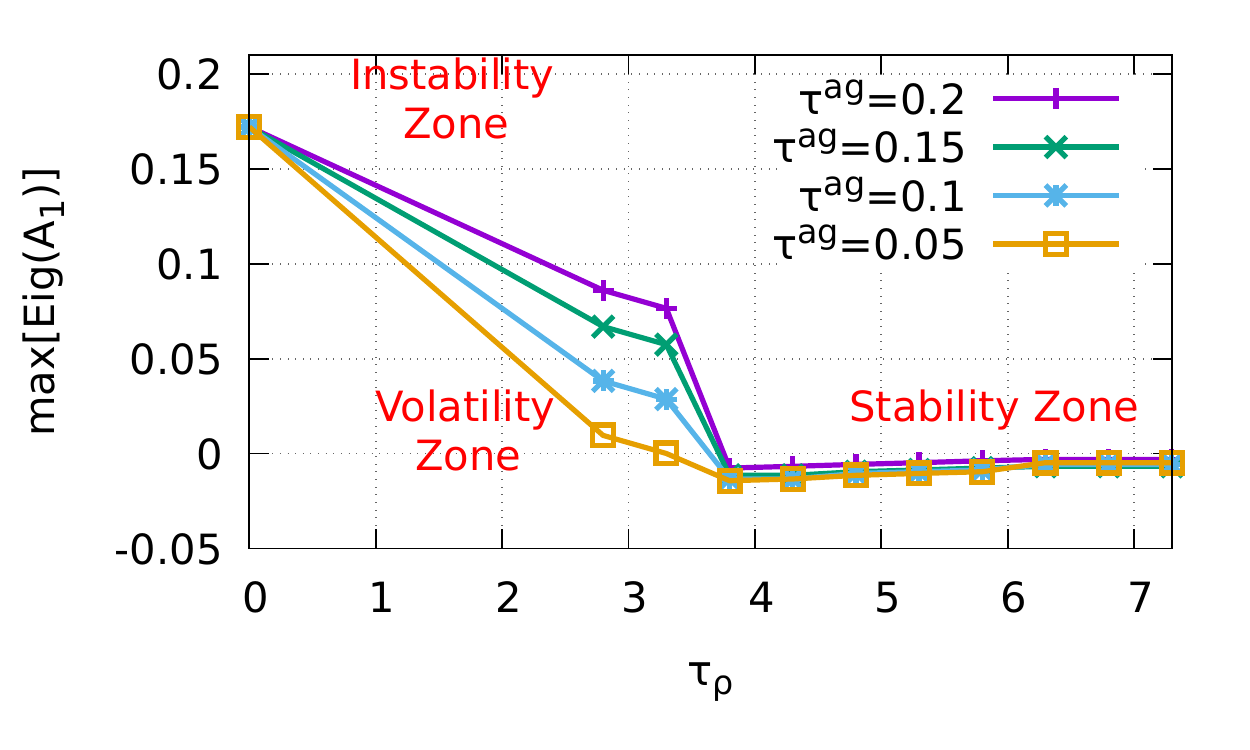}
  \includegraphics[width=0.33\textwidth]{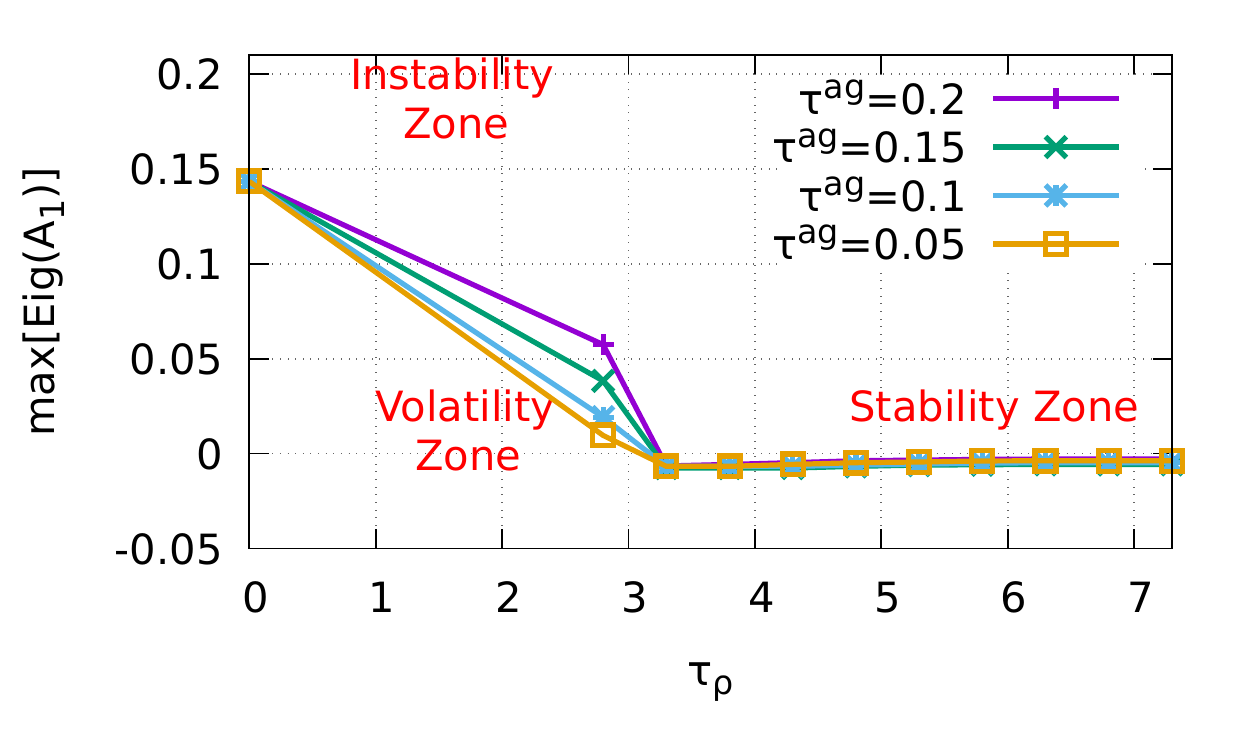} 
\end{tabular}
\caption{Market Stability Performance when (a) $\kappa_{1} = \kappa_{2} =0$ (left), (b) $\kappa_{1} = \kappa_{2} =0.02$ (middle), (c) $\kappa_{1} = \kappa_{2} =0.05$ (right)[Rawlsian]}
\end{figure*}

\begin{figure*}[htb]
\centering
 \begin{tabular}{@{}cc@{}}
  \includegraphics[width=0.33\textwidth]{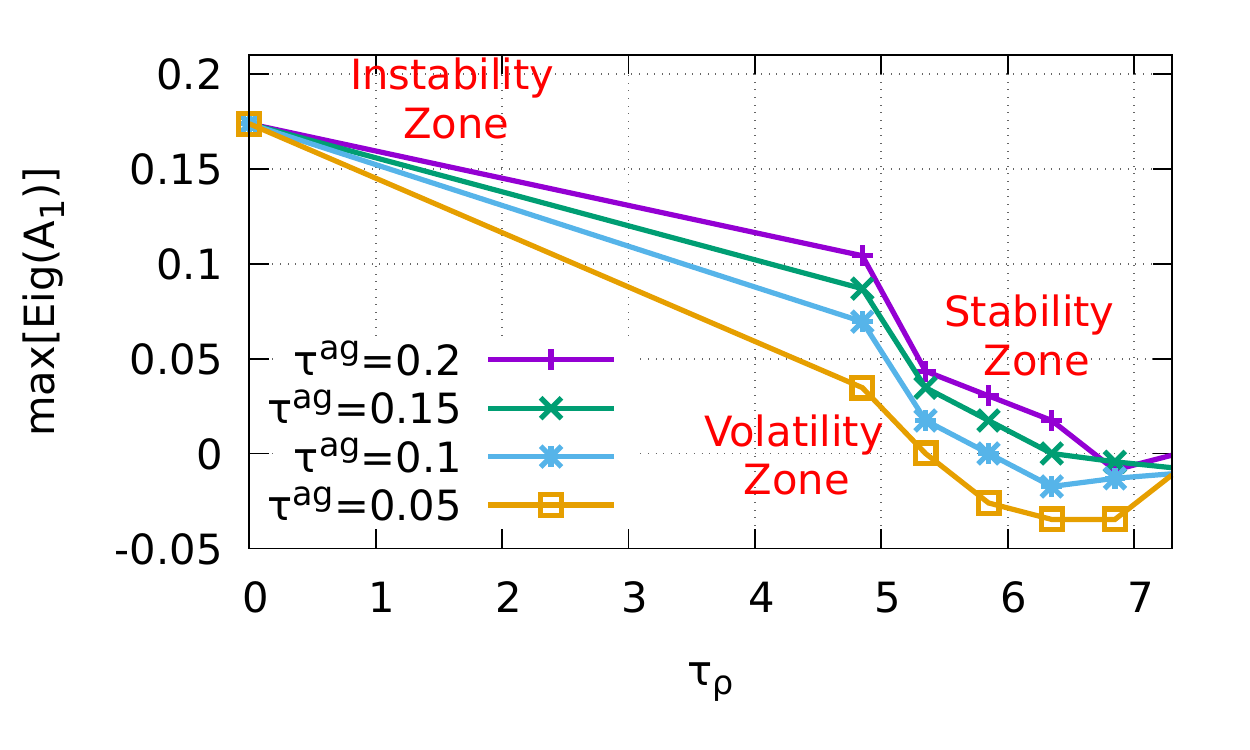} &
  \includegraphics[width=0.33\textwidth]{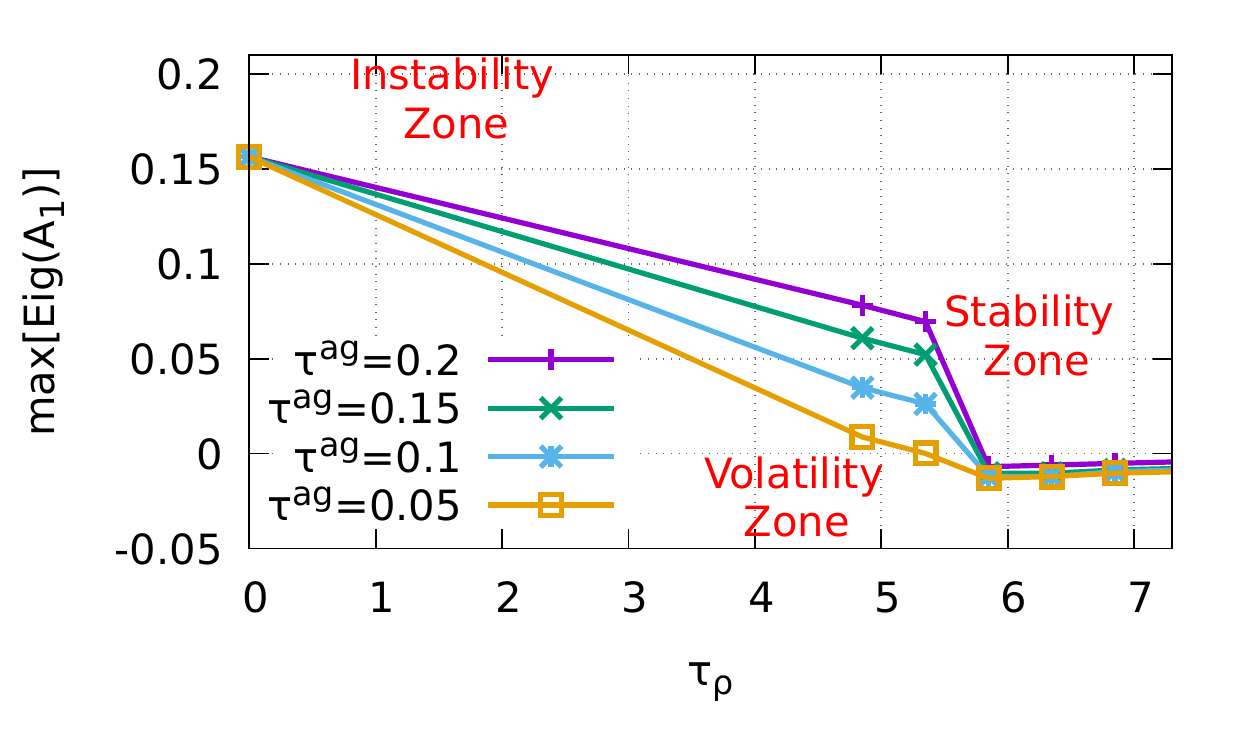}
  \includegraphics[width=0.33\textwidth]{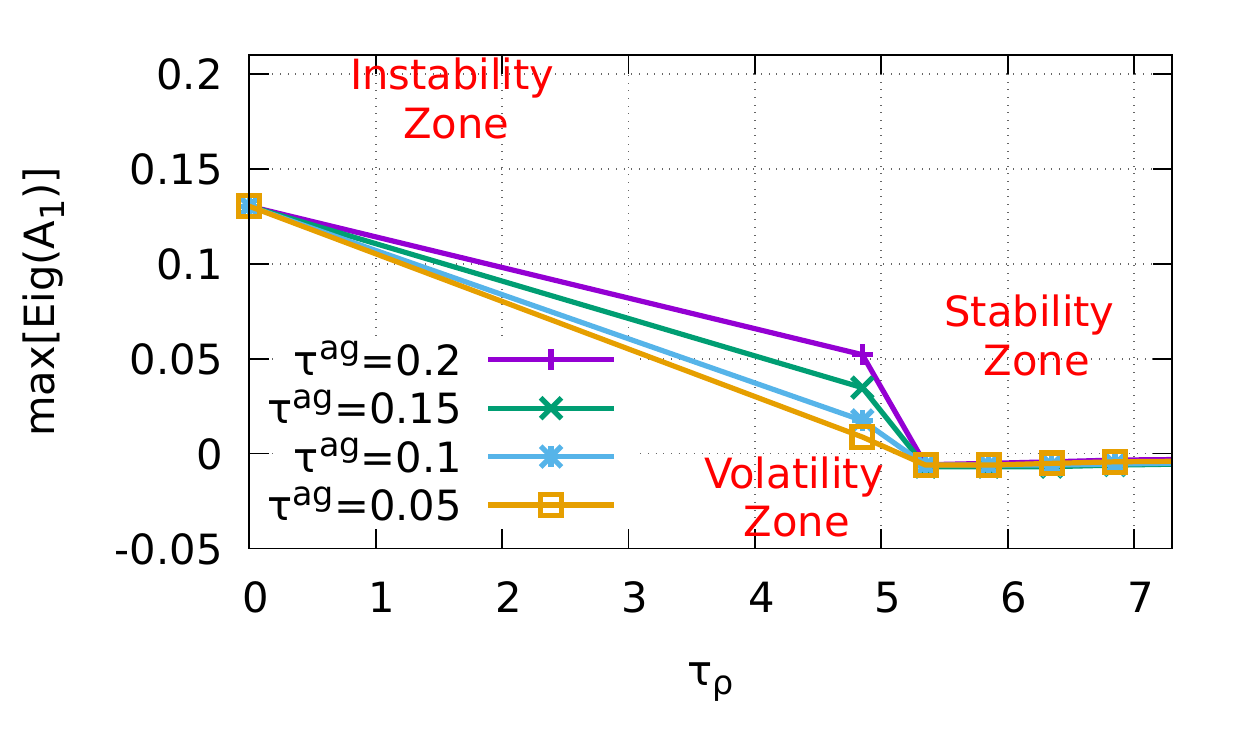} 
\end{tabular}
\caption{Market Stability Performance when (a) $\kappa_{1} = \kappa_{2} =0$ (left), (b) $\kappa_{1} = \kappa_{2} =0.02$ (middle), (c) $\kappa_{1} = \kappa_{2} =0.05$ (right)[Egalitarian]}
\end{figure*}
% -*- ispell-local-dictionary: "american"; TeX-master: "cloudshare.tex"; -*-
%
\section{Related Work}
We give an overview of efforts related to ours and highlight the relevant differences.
Works on hybrid clouds \cite{zhang2014proactive, shifrin2013optimal} are related
as they allow private (or smaller-scale) clouds
to outsource their requests to large-scale public providers. However, since that can potentially be costly
for a small-scale provider, our work differs in that it focuses on a sharing framework, while minimizing cost of using public clouds.

Earlier efforts also study the competition and cooperation within a federated cloud.
For instance, authors in \cite{goiri2012economic,hadji2015mathematical} characterize the cloud federation to help cloud providers maximize their profits via dynamic pricing models.
Earlier efforts \cite{truong2014novel,chen2017workload,cohen2014pricing,moon2010sla} also study the competition and cooperation among cloud providers, but assume that each cloud provider has sufficient resources to serve all users' requests, while \cite{chen2017workload} incorporates a penalty function to address the service delay penalty.
Authors in \cite{niyato2011resource} propose a hierarchical cooperative game theoretic model for better resources integration and achieving a higher profit in the federation.
Similarly to our work, \cite{mashayekhy2015cloud} studies a federation formation game but assumes that cloud providers share everything with others, while \cite{hassan2014cooperative} adopts cooperative game theoretic approaches to model a cloud federation and study the motivation for cloud providers to participate in a federation.

Another line of work focuses on designing sharing policies in the federation to obtain higher profit.
  For instance, \cite{wangenabling} proposes a decentralized cloud platform \emph{SpotCloud} \cite{spotcloud}, a real-world system allowing customers or SCs to sell idle compute resources at specified prices, and presents a resource pricing scheme (resulting from a repeated seller game) plus an optimal resource provisioning algorithm.
\cite{zhuang2014decentralizing} employs various cooperation strategies under varying workloads, to reduce the request rejection rate (i.e., the efficiency metric in \cite{zhuang2014decentralizing}).
Another effort \cite{toosi2011resource} combines resource outsourcing and rejection of less profitable requests in order to increase resource utilization and profit.
\cite{wen2016cost} proposes to efficiently deploy distributed applications on federated clouds by considering security requirements, the cost of computing power, data storage and inter-cloud communication.
\cite{babaoglu2012design} groups resources of various SCs into computational units, in order to serve customers' requests.
\cite{samaan2014novel} proposes to incorporate both historical and expected future revenue into VM sharing decisions in order to maximize an SC's profit.
%

%%%\emph{However, most of these efforts do not study the potential performance degradation received by each SC through participating in the cooperative, which is a significant factor for a SC to participate in the cooperative or not to; instead, they study the performance achievement of the whole cooperative.
%
%%%In contrast, our proposed effort is focused on achieving required
%%%performance characteristics of each SC while incentivizing them to participate in the cooperative, where
%%%some might be heavily loaded and others might have excess resources.

\smallskip\noindent\textbf{Differences and Drawbacks.}
Our work is a necessarily important theoretical extension of a very recent analytical work in \cite{lin2017performance} that was the first of its kind in the analysis of small cloud markets. There, the authors considered consequences of performance (i.e., queueing theory) driven non-cooperative game-theoretic (with no SC willing to share its utility and capacity information with others, i.e., an incomplete information game-theoretic setting) resource sharing on the resulting performance delivered to customers at \emph{static market equilibrium}, something \emph{not considered by any of the above-mentioned efforts.}
However,  \cite{lin2017performance} does not consider the important problem of analyzing equilibrium stability under variations in SC resource availability, in a non-cooperative game-theoretic SC environment. \emph{Without showing the existence of a stable SC market, one, based on the existing results showing the existence of a market equilibrium, cannot not say much regarding the sustainability of SC markets in the future.} A characterization of this scenario is an important contribution of this work.  A major difference of our work with the one in \cite{lin2017performance}, is the lack of a queuing-driven performance model to reduce the equilibrium search space. However, our work is orthogonal in the sense that, given the existence of (efficient) market equilibria, we investigate whether such a state is sustainable in the long run.

%Our work differs from previous efforts in that we explicitly consider consequences of resource sharing on the resulting performance delivered to customers. 
%%
%In contrast, none of the above efforts explicitly model the system performance under the considered resource sharing environment. 
%%
%They either assume that resources can be reclaimed (when needed), thus resulting in lack of reliability of shared resources, or they assume that an analytical performance characterization is possible (but do not propose a solution to estimate it).
%%
%Such an analytical characterization is an important contribution of our work.
%%
%To the best of our knowledge, this is the first work addressing the explicit interactions between performance model and economic model. 
%%
%Moreover, unlike previous efforts that adopt the cooperative game theoretic approach, our work studying the non-cooperative game is more practical since no SC would be willing to share its utility and capacity information with others.
%
%%assume the existence of the cloud federation and largely focus on designing sharing policies in order to maximize the profit of individual SCs, (2) most of them do not consider the risk of resource under-provision due to sharing too much resources - \emph{we focus on studying economic trade-off between the potential benefits (in terms of profit) and cost (in terms of performance degradation) for individual SCs, which are significant contributing factors in incentivizing SCs to participate in a cooperative}.
%

%\input{discuss}
\section{Conclusion and Future Work}
In this paper, we addressed the problem of effective resource sharing between small clouds (SCs). We modeled the problem as an efficient supply-demand market design task consisting of (i) autonomous SCs, (ii) their customers, and (iii) a regulator, as the market stakeholders. 
We first showed that a welfare allocation policy for the stakeholders by the regulator maximizes utilitarian social welfare at the static market equilibrium and results in the best/most efficient state at which the SC markets could operate. Fortunately, courtesy \emph{Arrow-Debreu} welfare theorems in welfare economics, this unique optimal operating point is also achieved in a distributed manner by the autonomous SCs in perfect price competition with one another, thereby guaranteeing no efficiency loss in a non-centralized market setting. 
However, the optimal market equilibrium point is prone to perturbations due to the dynamic nature of the SC market, thereby potentially leading to market disequilibrium. In this context, we designed a dynamic market mechanism based on Arrow and Hurwicz's disequilibrium process that uses the gradient play technique in game theory to converge upon the optimal static market efficient equilibrium from a disequilibrium state caused due to supply-demand perturbations, and results in market stability. We illustrated the stability and sustainability of dynamic SC markets and the high speed with which such markets converge to stable equilibria, through numerical experiments. 

\noindent \emph{A comment on Heterogenous VMs} - In out work, we have modeled a homogenous VM case. In practice, each cloud provider offers heterogeneous VM profiles (e.g., memory-optimized, CPU-optimized, or GPU-enabled), which reserve hardware resources on pre-specified machine pools shared by multiple VMs \cite{awstype}. However, many cloud providers, such as Amazon LightSail, DigitalOcean, and Linode, offer VM configurations with very similar specifications (e.g., \$10/month instances from Linode, DigitalOcean, and Amazon Lightsail currently provide 1 CPU core, 30 GB SSD, 2 TB data transfer/month, 1 or 2 GB of RAM). We believe that it is very likely that SCs would negotiate the sharing policies for each VM profile separately, given that these profiles correspond to different prices and capacities at each SC. In this case, our model of homogeneous resources can be applied repeatedly to each VM profile. Sharing policies for hardware resources (rather than VM profiles) would require the introduction of scheduling and packing algorithms within our performance model, which is beyond the scope of this work.

As part of future work, we plan to design provably fast distributed algorithms to allow markets to roll back to efficient equilibria when perturbed from an equilibrium state, and study dynamic SC markets under (i) a setting of imperfect competition between SCs, and (ii) a coalitional market setting where SCs have the capability to collude with one another.

\begingroup
%\setstretch{0.8}
\bibliographystyle{IEEEtran}
\bibliography{IEEEabrv,cloudshare,alluvion}
\endgroup
\newpage
\section{Appendix}
In this section, we provide the proofs for Theorems 4.1-4.3. 

%\noindent \textbf{Table Descriptions.} We state the SC and customer parameters through Tables 1 and 2 respectively as follows. 
%\begin{table}[h!]
%  \begin{center}
%    \caption{Parameters of Cost Functions for SCs}
%    \label{tab:table1}
%    %\hline
%    \begin{tabular}{c|c|c|c|c|c}
%      SC\# &$VM^{\min}$ & $VM^{\max}$ & $\tau_{SC}$ & $\beta$ & $\alpha$\\
%      \hline
%      SC1 & 0 & 200 & 0.6 & -0.3 & 90\\
%      SC2 & 0 & 200 & 0.2 & -0.6 & 102\\
%      SC3 & 0 & 250 & 0.6 & -0.25 & 80\\
%      SC4 & 0 & 250 & 0.6 & -0.25 & 80\\
%      SC5 & 0 & 200 & 0.2 & - 0.01 & 20\\
%     \hline
%    \end{tabular}
%  \end{center}
%\end{table}

%\begin{table}[h!]
%  \begin{center}
%    \caption{Parameters of Utility Functions for Customers}
%    \label{tab:table1}
%    %\hline
%    \begin{tabular}{c|c|c|c|c|c}
%      Customer(C)\# &$VM^{\min}$ & $VM^{\max}$ & $\tau_{C}$ & $\beta$ & $\alpha$\\
%      \hline
%      C1, C2 & 60 & 100 & 0.1 & -0.5 & 168\\
%      C3, C4 & 60 & 100 & 0.1 & -0.15 & 140\\
%      C5, C6 & 70 & 80 & 0.2 & -0.35 & 140\\
%      C7, C8 & 20 & 60 & 0.2 & -0.2 & 100\\
%      C9, C10 & 30 & 60 & 0.2 & -0.3 & 120\\
%      C11, C12 & 20 & 40 & 0.2 & -0.1 & 125\\
%      C13, C14, C15 & 30 & 60 & 0.2 & -0.5 & 135\\
%     \hline
%    \end{tabular}
%  \end{center}
%\end{table}

\noindent \textbf{Theorem Proofs.} We now state the theorem proofs below. 

\emph{Proof of Theorem 4.1.} The equilibrium $(x_{1}^{*}, x_{2}^{*})$ when setting $\kappa_{j}^{1}, \kappa_{j}^{2}$ to zero, is a solution of the following.
\begin{subequations}
\begin{align}
\rho_{i}^{*}  - \beta_{i}^{r}vm_{i}^{r*} - \alpha_{i}^{r}& = 0,\, \forall i\in SDC. \\
\rho_{i}^{*}  - \beta_{i}^{b}vm_{i}^{b*} - \alpha_{i}^{b}& = 0,\, \forall i\in SDC. \\
\rho_{i}^{*}  - \beta_{i}^{pc}vm_{i}^{pc*} - \alpha_{i}^{pc}& = 0,\, \forall i\in SDC. \\
\beta_{i}^{type}vm_{i}^{type} + \alpha_{i}^{type} - \rho_{i}^{*} & = 0,\, \forall i\in C,\,type \in \{e,c,s, ag\}.\\
\sum_{j\in C_{i}}vm_{j}^{ag}(1 - \kappa_{j}^{1} -\kappa_{j}^{2}) & = (vm_{i}^{r} + vm_{i}^{b} + vm_{i}^{pc}),\,\forall i\in SDC.
\end{align}
\end{subequations}
Using Theorem 3 in \cite{wangr}, strong duality implies that equilibrium $(x_{1}^{*}, x_{2}^{*}$ exists is identical to the solution of the KKT conditions in (10a)-(10e). It can be seen that (25a) follows by replacing the cost function for SDCs  in (2)-(4) in (10a). Similarly, (25b) follows by replacing the utility function of SDC customers in (5)-(8) in (10d). Furthermore (25c) is identical to (10e). Thus, $(x_{1}^{*}, x_{2}^{*}$ is identical to the equilibrium in (10a)-(10e). Thus, we proved Theorem 4.1. $\blacksquare$. \\

\emph{Proof of Theorem 4.2.} Since strong duality holds, it follows from Theorem 4.1 that equilibrium $(x_{1}^{*}, x_{2}^{*}\in E$ exists. We first prove the stability of this equilibrium point and then proceed to its asymptotic stability. Differentiating the positive definite Lyapunov function $V(y_{1}, y_{2}) = y_{1}^{T}P_{1}y_{1} + y_{2}^{T}P_{2}y_{2}$, with respect to time where $y_{1} = x_{1} - x_{1}^{*}$ and $y_{2} = x_{2} - x_{2}^{*}$, and by using the non-expansive property of the projection operation, we have 
{\small
\begin{equation}
\bar{V(y_{1}, y_{2})} \le y_{1}^{T}(P_{1}A_{1} + A_{1}^{T}P_{1})y_{1} + y_{1}^{T}P_{1}A_{2}y_{2} + y_{2}A_{2}^{T}P_{1}y_{1}
\end{equation}}
If $A_{1}$ is Hurwitz, for any $Q > 0$, there exists a positive definite matrix $P_{1}$ such that $P_{1}A_{1} + A_{1}^{T}P_{1} = -Q$. Let $\lambda_{\min}(Q)$ denote the minimum eigenvalue of $Q$. Since $P_{2}$ is a symmetric positive definite matrix with a set $n$ orthogonal, real, and non-zero eigenvectors $x_{1},....,x_{n}$, can be written as 
\[P_{2} = \sum_{i=1}^{n}\lambda_{i}x_{i}x_{i}^{T},\]
where $\lambda_{i} > 0$ is the eigenvalue corresponding to $x_{i}$. We can expand the vector $VM^{\max}$ using the orthogonal vector  $w_{i}$ as 
\begin{equation}
VM^{max^{T}}[1]_{n\times n}P_{2}y_{2} \ge \lambda_{\min}(P_{2})\psi_{\min}||y_{2}||_{2},
\end{equation}
where $\psi_{\min} = \min(\psi_{i}),\forall i = 1,..., n$. Now let 
\[\beta \ge ||P_{1}A_{2} + R^{T}[1]_{n\times n}P_{2}||_{2}.\]
Using (26) and (27), we obtain 
\begin{eqnarray*}
\bar{V(y_{1}, y_{2})} &\le& -\lambda(Q)\left(||y_{1}||_{2} - \frac{\beta}{\lambda_{\min}(Q)}||y_{2}||_{2}\right)^{2}\\
&-& ||y_{2}||\left(2\lambda_{\min}(P_{2}\psi_{\min} - \frac{\beta^{2}}{\lambda_{\min}(Q)}||y_{2}||\right).
\end{eqnarray*}
For all $\Omega_{\max} \subsetneq D$, it follows that for all solutions beginning in $\Omega_{max}$, $V \le 0$. Hence, the equilibrium is stable and $\Omega_{\max}$ is the region of attraction. 

Since the initial conditions start in $\Omega_{\Delta}$ and the latter is a strict subset of $D_{\Delta}$,  $y_{2}$ cannot be equal to $2\lambda_{\min}(P_{2})\psi_{\min}\frac{\lambda_{\min}(Q)}{\beta^{2}}$. This in turn implies that $(||y_{1}||,||y_{2}|| = (0,0)$ is the only invariant set. Hence, all solutions starting in $\Omega_{\Delta}$ converge to the equilibrium point $(x_{1}, x_{2}) = (x_{1}^{*}, x_{2}^{*})$. Thus, we proved Theorem 4.2. $\blacksquare$\\

\emph{Proof of Theorem 4.3.} Differentiating the Lyapunov function $V(y_{1}, y_{2})$ along the trajectories of (16), we get 
{\small
\begin{equation}
\bar{V(y_{1}, y_{2})} \le -a_{\Delta}\left(||y_{1}|| - \frac{\beta}{a_{\Delta}}||y_{2}||\right)^{2} - ||y_{2}||\left(e - \frac{\beta^{2}}{a_{\Delta}}||y_{2}||\right),
\end{equation}}
where $a_{\Delta} = \lambda_{\min}(Q) - 2||P_{1}||\pi_{SDC} + 2||P_{1}||\pi_{C}$, and $e = 2\lambda_{\min}(P_{2})\psi_{\min}$. 

From (24) it follows that $a_{\Delta} > 0$. Therefore, (25) implies that for all $\Omega_{c_{max}} \subsetneq D_{\Delta}$, for all solutions beginning in $\Omega_{\Delta}$, $\bar{V} \le 0$. Hence, the market equilibrium state is stable, and $\Omega_{\Delta}$ is the region of attraction. 

The asymptotic stability of the perturbed market can be shown via the following argument: since the initial conditions start in $\Omega_{\Delta}$ and the latter is a strict subset of $D_{\Delta}$,  $y_{2}$ cannot be equal to $2\lambda_{\min}(P_{2})\psi_{\min}\frac{\lambda_{\min}(Q)}{\beta^{2}}$. This in turn implies that $(||y_{1}||,||y_{2}|| = (0,0)$ is the only invariant set. Hence, all solutions starting in $\Omega_{\Delta}$ converge to the equilibrium point $(x_{1}, x_{2}) = (x_{1}^{*}, x_{2}^{*})$. Thus, we have proved Theorem 4.3. $\blacksquare$

\end{document}